\documentclass{jfm}
\usepackage{graphicx}
\usepackage{epstopdf, epsfig}
\usepackage{float}
\usepackage{amsfonts}
\usepackage{bm}
\usepackage{hyperref}
\usepackage{mathtools}
\usepackage{color}

\usepackage[utf8]{inputenc}
\usepackage[english]{babel}

\usepackage[dvipsnames]{xcolor}
\colorlet{LightRubineRed}{RubineRed!70!}
\colorlet{Mycolor1}{green!10!orange!90!}
\definecolor{Mycolor2}{HTML}{00F9DE}

\shorttitle{Comparing cascade rates in turbulence}
\shortauthor{H. Yao, M. Schnaubelt, A.S. Szalay, T. Zaki and C. Meneveau}

\title{Comparing local energy cascade rates in isotropic turbulence using structure function and filtering formulations}

\author{H. Yao\aff{1}\corresp{\email{hyao12@jhu.edu}},
  M. Schnaubelt\aff{2},
  A. Szalay\aff{2}, T. Zaki\aff{1} \and 
  C. Meneveau\aff{1}}

\affiliation{\aff{1}Department of Mechanical Engineering \& IDIES, Johns Hopkins University
\aff{2}Department of Physics and Astronomy \& IDIES, Johns Hopkins University
}

\begin{document}

\maketitle

\begin{abstract}
Two common definitions of the spatially local rate of kinetic energy cascade at some scale $\ell$ in turbulent flows are (i) the cubic velocity difference term appearing in the generalized Kolmogorov-Hill equation (GKHE) (structure function approach), 
and (ii) the subfilter-scale energy flux term in the transport equation for subgrid-scale kinetic energy (filtering approach). We perform a comparative study of both quantities based on direct numerical simulation data of isotropic turbulence at Taylor-scale Reynolds number of 1250. While observations of negative subfilter-scale energy flux (backscatter) have in the past led to debates regarding interpretation and relevance of such observations, we argue that the interpretation of the local structure function-based cascade rate definition is unambiguous since it arises from a divergence term in scale space.  Conditional averaging is used to explore the relationship between the local cascade rate and the local filtered viscous dissipation rate as well as filtered velocity gradient tensor properties such as its invariants. We find statistically robust evidence of inverse cascade when both the large-scale rotation rate is strong and the large-scale strain rate is weak. Even stronger net inverse cascading is observed in the ``vortex compression'' $R>0$, $Q>0$ quadrant where $R$ and $Q$ are  velocity gradient invariants. Qualitatively similar, but quantitatively much weaker trends are observed for the conditionally averaged subfilter scale energy flux.  Flow visualizations show consistent trends, namely that spatially the inverse cascade events appear to be located within  large-scale vortices, specifically in subregions when $R$ is large.
\end{abstract}


\newpage
\section{Introduction}
\label{sec:intro}

The classic description of the energy cascade in turbulence postulates that kinetic energy originates from forcing large-scale eddies, is subsequently transferred to smaller-scale eddies (forward cascade), and is eventually dissipated due to viscous effects \citep{richardson1922weather,kolmogorov1941local}. In a statistical sense, the sign and magnitude of third-order moments of velocity increments confirm this general direction of the energy cascade, as described by the $4/5$ law governing the global average of the third-order longitudinal velocity increment 
\citep{kolmogorov1941local,frisch1995turbulence},
$
\langle \delta u_L(\ell)^3 \rangle \equiv \langle  ( [{\bf u}({\bf x}+{\bm \ell}) - {\bf u}({\bf x})]\cdot  {{\bm \ell} / \ell }  )^3 \rangle 
=- \frac{4}{5} \, \ell \, \langle \epsilon \rangle$, where $\langle ... \rangle$ denotes global averaging, $\delta u_L(\ell)$ is the longitudinal  velocity increment and $\epsilon$ is the viscous dissipation rate, while the displacement $\ell=|{\bm \ell}|$ is assumed to be well inside the inertial range of turbulence. 
In this sense, the quantity $-\frac{5}{4}\langle \delta u_L(\ell)^3 \rangle/\ell$ is often interpreted as a measure of the energy flux going from scales larger than $\ell$ to all smaller scales. Because turbulence is known to be highly intermittent in space and time \citep{kolmogorov1962refinement,frisch1995turbulence,meneveau1991multifractal} there has also been much interest in characterizing the local properties of the energy cascade, i.e. the fluctuations of the energy flux before averaging. However, without statistical averaging, the $4/5$-law is less meaningful, e.g., the quantity  $-\frac{5}{4}  \delta u_L^3  /\ell$ cannot simply be interpreted as an energy flux locally in space and time. To enable such interpretation, it is necessary to consider explicit angular averaging over all possible directions of the vector $\bm \ell$. Such formulations have been developed in prior works by  \cite{duchon2000inertial}, \cite{eyink2002local} and   \cite{hill2001equations, Hill_2002a}. \cite{duchon2000inertial} and \cite{eyink2002local} use such equations to study the energy cascade and energy dissipation in the limit of zero viscosity.  A review about extensions to the classic Kolmogorov equation is presented by \cite{dubrulle2019beyond}, specifically focusing on the \cite{duchon2000inertial} local formulation.

\cite{hill2001equations, Hill_2002a} developed a local version of the Kolmogorov equation in which the reference position ${\bf x}$ is symmetrically located halfway between the two points ${\bf x}+{\bf r}/2$ and ${\bf x}-{\bf r}/2$ separated by ${\bf r}$ over which the velocity increment is computed. This equation, which we shall denote as the generalized Kolmogorov-Hill equation (GKHE), describes the evolution of the second-order (squared) velocity difference, a measure of energy content of all scales smaller than $|\bm r|$  at a specific physical position $\bf x$. As will be reviewed in \S \ref{sec:equations}, scale-space integration over ${\bf r}$ of the GKHE up to some scale $\ell$ in the inertial range and without additional statistical averaging  provides a localized description of the energy cascade process.  The GKHE also includes  effects of viscous dissipation, viscous diffusion,  advection, and pressure. A number of prior works have studied various versions of the GKH equation. For isotropic turbulence, \cite{yasuda2018spatio} quantified the variability of the energy flux that arises in this equation,  while \cite{carbone2020vortex} considered a definition of mean energy flux approximated based on a solenoidal filtered velocity increments and examined its connections to average vortex and strain stretching rates. Besides applications to isotropic homogeneous flow,  numerous studies have investigated the application of the statistically averaged GKHE to spatially non-homogeneous flows. For instance, in wall boundedf lows, researchers have explored the energy cascade using a Reynolds decomposition to isolate effects of mean shear and non-homogeneity \citep{
Antonia_et_al_2000,danaila2001turbulent,Danaila_et_al_2004,Danaila_et_al_2012,
marati2004energy, cimarelli2013paths}. Investigations have also studied the energy cascade rates in boundary layer bypass transition \citep{yao2022analysis}, and flow separation \citep{mollicone2018turbulence}. Furthermore, specific attention has been given to the study of inverse cascade in wake flows \citep{gomes2015energy, portela2017turbulence} and at turbulent/non-turbulent interfaces \citep{zhou2020energy, cimarelli2021spatially,yao_papadakis_2023}.

The notion of transfer, or flux, of kinetic energy across length-scales is of particular practical interest  also in the context of large eddy simulation (LES). There the rate of energy cascade is commonly referred to as the subgrid or subfilter-scale (SGS, SFS) rate of dissipation. It is defined as the contraction between the subgrid stress tensor and the filtered strain-rate tensor and arises  as a source term in the transport equation for subgrid/subfilter-scale kinetic energy \citep{piomelli1991subgrid,meneveau2000scale}. This quantity characterizes the energy transfers between the resolved scale and the residual scale within the inertial range, which is also a local property \citep{eyink2009localness}. The SGS dissipation is highly intermittent \citep{cerutti1998intermittency}, and can be both positive and negative locally, but on average, energy is known to be transferred from large scales to the residual scales (forward cascade). There is considerable literature on the subject starting from the seminal papers by \cite{lilly1967representation}, \cite{leonard1975energy} and \cite{piomelli1991subgrid}. Some reviews include \cite{meneveau2000scale,meneveau2010turbulence,moser2021statistical}.

Without averaging, it has been a common observation that the SGS/SFS dissipation can be negative which has often been interpreted as indicative of local inverse cascading of kinetic energy, i.e.,  energy transfering from  small  to large scales of motion (``backscatter'' \citep{piomelli1991subgrid}).   \cite{borue1998local} noted that the forward cascade occurs predominantly in regions characterized by strong straining, where the magnitude of negative skewness of the strain tensor and vortex stretching are large. Conversely, backscatter was observed in regions with strong rotation. The relationship between SGS dissipation and stress topology and stress strain alignment geometry is discussed and measured based on 3D PIV measurements by   \cite{tao2002statistical}. In a more recent study,  \cite{ballouz2018tensor} investigated the SGS tensor by considering the relative alignment of the filtered shear stress and strain tensors. They found that the energy cascade efficiency is quite low, a trend they attributed to energy being transfered largely between positions in physical space. Quantitatively, in expressing the subgrid-stress tensor as a superposition of all smaller scale Gaussian-filtered velocity gradients,  \cite{johnson2020energy,johnson2021role} was able to isolate the relative contributions of small-scale strain self-stretching and vortex stretching finding both to be important. 

It has been questioned whether it is the local quantity $-\tau_{ij}\tilde{S}_{ij}$ (where $\tau_{ij}$ and $\tilde{S}_{ij}$ are the subgrid-scale stress and resolved strain-rate tensors, respectively), or the work done by the SGS/SFS force, $\tilde{u}_i\partial_j\tau_{ij}$ (where $\tilde{u}_i$ is the resolved velocity), that should be the genuine definition of local energy cascade rate. For instance \cite{kerr1996small} use the latter in their study of correlations of cascade rate and vorticity, and more recently \cite{vela2021entropy} uses both quantities in their analysis. Moreover the SGS force plays a central role for optimal LES modeling \citep{langford1999optimal}. The SGS force is invariant to divergence-free tensor fields which therefore do not affect the large-scale dynamics but addition of such a tensor field to $\tau_{ij}$ can certainly affect the usual definition of subgrid-scale dissipation $-\tau_{ij}\tilde{S}_{ij}$.  By re-expressing the SGS stress and dissipation terms using an optimization procedure, \cite{vela2022subgrid} provided arguments that the often observed  backscatter does not actually contribute to the energy cascade between scales but rather to the energy flux in the physical space, also suggesting that backscatter does not need to be explicitly modeled in LES. 

As can be seen from this partial summary of the literature on backscatter and inverse cascade in the LES filtering approach, no consensus has been reached regarding the possible importance and physical interpretation of local backscatter using the definition based on the inner product of the subgrid stress and filtered strain-rate tensors. Also, the question of inverse cascade has not received much attention from the point of view of the local versions of the Kolmogorov equation in the structure function approach. Therefore, in this paper we first revisit the generalized local structure function formulation  (\S \ref{sec:equationssf}). We argue that in this formulation the term responsible for the energy cascade can be unambiguously interpreted as a flux of kinetic energy between scales since it appears inside a divergence in scale space. In this sense it differs from the filtering formulation used in LES (reviewed in \S \ref{sec:equationsLES}) in which typically a fixed filter scale is used and no change in scales is considered, thus making the concept of a ``flux in scale space'' less clearly defined and open to various interpretations.   

With the definition of local cascade rate or energy flux clarified for the structure function approach, we perform a comparative study of both the structure function and the filtering approaches' energy flux terms in a relatively high Reynolds number DNS database of forced isotropic turbulence at a Taylor-scale Reynolds number of 1,250. The data analysis is greatly facilitated by the availability of these data in a new version of the Johns Hopkins Turbulence Database (JHTDB) System, in which python notebooks access the data directly (see appendix \ref{appA}). The comparisons involve various statistical properties of the energy flux. First, in \S \ref{sec:compstats} we provide comparisons of both quantities by means of simple statistical measures such as their mean values, joint probability density distributions and correlation coefficients, comparing both the two definitions of kinetic energy and kinetic energy cascade rate or flux. We then comparatively examine conditional averages based on the local molecular dissipation rate averaged over a ball of size $\ell$, specifically re-examining the Kolmogorov refined similarity hypothesis (KRSH) in \S \ref{sec:compKRSH}. Then, in \S \ref{sec:compvelgrad}, we present comparative conditional averages of kinetic energy flux based on properties of the large-scale velocity gradient field such as the strain and rotation magnitudes, and the $Q$ and $R$ invariants.  Particular attention is placed on events of local negative energy flux and whether or not such events can be considered to be of statistical significance. Overall conclusions are presented in \S \ref{sec:conclusions}.

\section{Local energy flux in the structure function and filtering approaches}
\label{sec:equations}

In this section, both the structure function based (GKHE) and filtering (LES)  energy equations are reviewed. We focus on the term representing energy cascade (energy flux) in each equation, and describe some of the prior efforts in the literature  relating the structure function and filtering approaches.  

\subsection{Energy cascade rate/flux in the generalized Kolmogorov-Hill equation}
\label{sec:equationssf}

\begin{figure}
 \centering
  \includegraphics[scale=0.3]{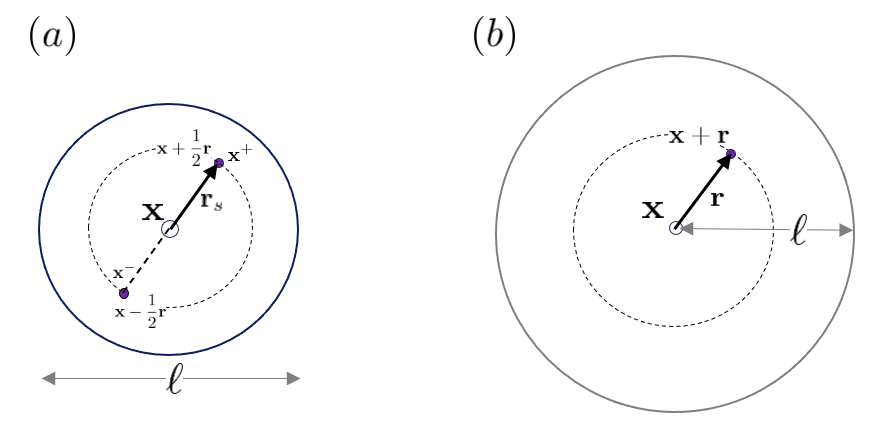}
    \caption{(a): Sketch showing local domain of integration over a ball of diameter $\ell$  used in the symmetric \cite{hill2002exact} structure function approach in which pairs of points separated by distances $r=2r_s$ up to $\ell $ are used. (b) shows integration up to a ball of radius $\ell$ in which pairs of points separated by distances $r$ up to $\ell $ are used as in the approach of \cite{duchon2000inertial}. For volume averaging, in (a) 3D integration over the vector ${\bf r}_s$ is performed at fixed ${\bf x}$ while in (a) 3D integration over the vector ${\bf r}$ is performed at fixed ${\bf x}$. For surface integrations, in (a) integration is done over the spherical surface of radius $\ell/2$ while in (b) is is done over a spherical surface of radius $\ell$.}
    \label{sketch}
\end{figure}

The GKHE is a generalized Karman-Horwath equation that is directly derived from the incompressible Navier-Stokes equations without any modeling. Before averaging, the instantaneous GKHE with no mean flow and neglecting the forcing term reads \citep{hill2001equations, hill2002exact}:
\begin{equation}
 \begin{aligned}
\frac{\partial \delta u _i^2}{\partial t} + u^*_{j}\frac{\partial \delta u _i^2}{\partial x_j}  = 
-\frac{\partial \delta u _j\delta u _i^2}{\partial r_j}-\frac{8}{\rho}\frac{\partial p^*\delta u _i}{\partial r_i} 
+\nu \frac{1}{2} \frac{\partial^2 \delta u _i  \delta u _i}{\partial x_j \partial x_j
}+ 
2\nu \frac{\partial^2 \delta u _i \delta u _i}{\partial r_j \partial r_j}
-
4\epsilon^*,
\label{ins_KHMH_noint}
\end{aligned}
\end{equation} where $\delta u_i = \delta u_i({\bf x},{\bf r}) =  u_i^+ - u_i^-$ is the velocity increment vector in the $i^{\textrm{th}}$ Cartesian direction over displacement vector ${\bm r}$. The superscripts $+$ and $-$ represent two points ${\bf x}+{\bf r}/2$ and ${\bf x}-{\bf r}/2$ in the physical domain that have a separation vector $r_i = x^+_i - x^-_i$ and middle point 
$x_i = (x^+_i + x^-_i)/2$ (see Fig. \ref{sketch} (a)). The superscript $*$ denotes the average value between two points, e.g., the two-point average dissipation is defined as $\epsilon^*({\bf x},{\bf r}) = (\epsilon^+ +\epsilon^-)/2$, and $\epsilon^\pm$ here is the ``pseudo-dissipation'' defined at every point as $\epsilon =\nu  ({\partial u_i}/{\partial x_j})^2$. Also, we will use ${\bf r}_s = {\bf r}/2$ to denote the radial coordinate vector  from the local ``origin'' ${\bf x}$. 

As remarked by  \cite{hill2001equations, hill2002exact} it is then instructive to apply integration over a sphere in ${\bf r}_s$-space up to a radius $\ell/2$, i.e. over a sphere of diameter  $\ell$.  The resulting equation is divided by the sphere volume  $V_\ell=\frac{4}{3}\pi( {\ell}/{2})^3$ and a factor 4, and Gauss' theorem is used for the ${\bf r}$-divergence terms (recalling that $\partial{\bf r}=2\partial{\bf r}_s$), yielding

\begin{equation}
 \begin{aligned}
\frac{1}{2 \,V_\ell}\iiint\limits_{V_{\ell}} \left(\frac{\partial \delta u _i^2/2}{\partial t} + u^*_{j} \frac{\partial \delta u _i^2/2}{\partial x_j} \right) d^3{\bf r}_s
=
-\frac{3}{4\ell}\frac{1}{S_\ell}\oint\limits_{S_{\ell}} \delta u _i^2 \delta u _j \hat{n}_j  dS
-
\frac{6}{\rho \,\ell}\frac{1}{S_\ell}\oint\limits_{S_{\ell}}p^*\delta u _j \hat{n}_j dS \\
+
\frac{\nu}{4}\frac{1}{V_\ell}\iiint\limits_{V_{\ell}}\left(\frac{1}{2} \frac{\partial^2 \delta u _i  \delta u _i}{\partial x_j \partial x_j 
}
+
2 \frac{\partial^2 \delta u _i \delta u _i}{\partial r_j \partial r_j}
\right)
d^3{\bf r}_s
-
 \frac{1}{V_\ell}\iiint\limits_{V_{\ell}}
\epsilon^*
d^3{\bf r}_s,
\end{aligned}
\label{ins_GKHE_noint}
\end{equation}
where $S_\ell$ represents the bounding sphere's surface of area $S_\ell = 4\pi(\ell/2)^2$ and $\hat{n}_j$ is the unit normal vector. 
Eq. \ref{ins_GKHE_noint}  suggests defining a structure-function based kinetic energy at scale $\ell$ according to
\begin{equation}
k_{{\rm sf},\ell}({\bf x},t) = \frac{1}{2 \,V_\ell}\iiint\limits_{V_{\ell}}  \frac{1}{2} \delta u _i^2({\bf x},{\bf r}) \, d^3{\bf r}_s,
\end{equation}
so that the first term in Eq. \ref{ins_GKHE_noint} corresponds to $\partial \, k_{{\rm sf},\ell}/\partial t$.
The 1/2 factor in front of the integral is justified since the volume integration over the entire sphere will double count the energy contained in $\delta u _i^2 = u_i^+ - u_i^-$. 
Equation \ref{ins_GKHE_noint} thus describes the transport of two-point, structure function energy $k_{{\rm sf},\ell}$, which represents energy within eddies with length scales up to  $\ell$ \citep{davidson2015turbulence} in both the length scale $\ell$ and physical position $\bf x$ spaces. 
The last term in equation \ref{ins_GKHE_noint} represents $r$-averaged rate of dissipation with the radius vector
 ${\bf r}_s = {\bf r}/2$ being integrated up to magnitude $\ell/2$,

\begin{equation}
\epsilon_\ell({\bf x},t) \equiv \frac{1}{V_\ell}\iiint\limits_{V_{\ell}}
 \epsilon^*({\bf x},{\bf r}) \,
d^3{\bf r}_s.
\end{equation}
As remarked by  \cite{hill2001equations, hill2002exact} this quantity corresponds directly to the spherical average of local dissipation at scale $\ell$ and plays a central role in the celebrated Kolmogorov Refined Similarity Hypothesis, KRSH \citep{kolmogorov1962refinement}.  

The local energy cascade rate in the inertial range at position ${\bf x}$ and time $t$ is defined as
\begin{equation}
\Phi_\ell({\bf x},t)  \equiv -\frac{3}{4\,\ell}\frac{1}{S_\ell}\oint\limits_{S_{\ell}} \delta u _i^2\,\delta u _j\,  \hat{n}_j \, dS = -\frac{3}{4\,\ell} \, [\delta u_i^2 \delta u_j \hat{n}_j ]_{S_\ell},
\label{phiflux}
\end{equation} where $[..]_{S_\ell}$ indicates area averaging over the sphere of diameter $\ell$. We note that in this definition, $\Phi_\ell({\bf x},t)$ represents the surface average of a flux that is defined positive if energy is flowing into the sphere in the ${\bf r}$-scale space. The position is fixed at ${\bf x}$ and thus the quantity $\Phi_\ell({\bf x},t)$  does not contain possible confounding spatial transport effects. 

In terms of the overall average of Eq. \ref{ins_GKHE_noint}, under the assumptions of homogeneous isotropic flow and statistical steady-state conditions, and for $\ell$ in the inertial range of turbulence, the unsteady, transport and viscous terms vanish. The pressure term is also zero due to isotropy and incompressibility. Therefore, Eq. \ref{ins_GKHE_noint} can be simplified and yields as expected 
\begin{equation}
\langle \Phi_\ell \rangle
=
\langle \epsilon_\ell \rangle=\langle \epsilon \rangle, 
\label{local_eq}
\end{equation} 
or equivalently $[\delta u_i^2 \delta u_j \hat{n}_j ]_{S_\ell}=-4/3\, 
\ell\, \langle \epsilon \rangle$, the 4/3-law \cite{frisch1995turbulence}.

In this paper the focus will be mainly on the flux term $\Phi_\ell$ with some attention also on the dissipation term $\epsilon_\ell$. Analysis of the time derivative, spatial advection terms and pressure terms is left for other ongoing studies. The viscous flux terms (in both spatial and scale spaces) are also not considered, since our present interest concerns the inertial range.

\subsection{Energy cascade rate/flux in the filtering approach}
\label{sec:equationsLES}

In this section, we review the transport equation of the subgrid-scale kinetic energy   \citep{germano1992turbulence} for $k_{{\rm sgs},\ell}  \equiv \frac{1}{2} \tau_{ii}$, where $\tau_{ij}=\widetilde{u_i u_j} - \tilde{u}_i \tilde{u}_j$ is the subgrid-scale stress tensor, where  the tilde symbol $(\sim)$ denotes spatial filtering of variables. The transort equation for $k_{{\rm sgs},\ell}$ reads 
\citep{germano1992turbulence}

\begin{equation}
 \begin{aligned}
\frac{\partial k_{{\rm sgs},\ell}}{\partial t} + \tilde{u}_j \frac{\partial   k_{{\rm sgs},\ell} }{\partial x_j}= 
& 
-\frac{1}{2}\frac{\partial}{\partial x_j}\left(\widetilde{u_i u_i u_j}- 2\tilde{u}_i\widetilde{u_i u_j}-\tilde{u}_j\widetilde{u_i u_i}-\tilde{u}_i \tilde{u}_i \tilde{u}_j\right)
 -\frac{\partial}{\partial x_j}\left( \widetilde{p u_j} - \tilde{p}\tilde{u}_j
\right) 
\\
&
+\frac{\partial}{\partial x_j}\left(
\nu\frac{\partial k_{{\rm sgs},\ell} }{\partial x_j}
\right) 
+
\nu
\frac{\partial \tilde{u}_i}{\partial x_j}
\frac{\partial \tilde{u}_i}{\partial x_j}
-\nu\widetilde{\frac{\partial u_i}{\partial x_j}\frac{\partial u_i}{\partial x_j}}
-\tau_{ij}\widetilde{S}_{ij} .
 \label{LES_energy}
\end{aligned}
\end{equation}
The last term  is called the subgrid-scale rate of dissipation at  position $({\bf x})$, and is often denoted as
\begin{equation}
 \Pi_\ell({\bf x},t) \equiv -\tau_{ij}\widetilde{S}_{ij}.
 \label{Pi_r}
\end{equation}
For filtering, in the present work we consider a spherical-shaped sharp top-hat filter in physical space with a diameter equal to $\ell$. Therefore, for any field variable $A({\bf x})$, we define the filtered variable as $\tilde{A}({\bf x}) = {V_\ell}^{-1}\iiint\limits_{V_{\ell}} A({\bf x}+{\bf r}_s) \, d^3{\bf r}_s $. Note that each term in Equation \ref{ins_GKHE_noint} and in Equation \ref{LES_energy} are thus evaluated at the same length scale. Terms in Equation \ref{LES_energy} can be compared directly to terms in Equation \ref{ins_GKHE_noint}, in particular the local dissipation terms are exactly the same, i.e, 

\begin{equation}
-\nu\widetilde{\frac{\partial u_i}{\partial x_j}\frac{\partial u_i}{\partial x_j}} = -\frac{1}{V_\ell}\iiint\limits_{V_{\ell}} \nu\frac{\partial u_i}{\partial x_j}\frac{\partial u_i}{\partial x_j} d^3{\bf r}_s = \epsilon_\ell({\bf x},t).
 \label{}
\end{equation}Again, for homogeneous steady state turbulence in the inertial range (neglecting viscous diffusion and resolved dissipation terms), upon averaging Eq. \ref{LES_energy} simplifies to
\begin{equation}
\langle \Pi_\ell \rangle
=
\langle \epsilon_\ell \rangle =\langle \epsilon \rangle
\label{local_eq_pi}
\end{equation} which is similar to Equation \ref{local_eq} and thus on average certainly both definitions of energy cascade rate/flux agree with each other, i.e., $\langle \Pi_\ell\rangle =  \langle \Phi_\ell \rangle$.

It is also of interest to compare the average value of the two definitions of kinetic energy used in both definitions of energy cascade rate/flux. In the inertial range of high Reynolds number turbulence, both $\langle k_{{\rm sf},\ell}\rangle$ and $\langle k_{{\rm sgs},\ell}\rangle$  can be evaluated based on the Kolmogorov $r^{2/3}$ law and $k^{-5/3}$ spectrum, respectively. The result is (see Appendix \ref{appB} for details), $\langle k_{{\rm sf},\ell} \rangle = 1.575 \, \langle \epsilon \rangle^{2/3} \ell^{2/3}$ and 
$\langle k_{{\rm sgs},\ell}\rangle = 1.217 \, \langle \epsilon \rangle^{2/3} \ell^{2/3}$. In other words, they are of similar order of magnitude but the SGS kinetic energy is slightly smaller. 

\subsection{Other relationships between structure function and filtering approaches}

In the present paper, we shall perform the data analysis and comparisons using the two approaches mentioned above (GKHE and filtering). However, it is useful at this stage to include some remarks regarding other structure function and energy definitions used in earlier works by \cite{vreman1994realizability}, \cite{constantin94onsager}, \cite{duchon2000inertial}, \cite{eyink2002local} and \cite{dubrulle2019beyond}. Those approaches typically focus on the structure function written at one of the endpoints instead of the midpoint.
\cite{duchon2000inertial} and \cite{dubrulle2019beyond} focus on the two-point correlation quantity  $C({\bf x},{\bf r}) = u_i({\bf x}) u_i({\bf x} + {\bf r})$ (see Fig. \ref{sketch} (b)). Local averaging over all values of ${\bf r}$ from ${\bf r}=0$ up to scale $|{\bf r}|=\ell$ at any given ${\bf x}$ then corresponds to the ``mixed'' energy quantity $u_i \tilde{u}_i/2$ (denoted as $E^{\ell}$ in \cite{dubrulle2019beyond}), and where the filtering is over a sphere of diameter  $2\ell$ so as to combine two points with separation distances up to $\ell$. The quantity $C({\bf x},{\bf r})$ combines filtered and unfiltered velocities and hence it is more difficult to interpret for comparisons of structure function and LES filtering approaches. In its transport equation,  
\cite{duchon2000inertial} show that a term similar to the third-order structure function term of Eq. \ref{phiflux} arises. However, in order for the structure function to correspond to scale $\ell$, one has to choose to integrate over a sphere of diameter $2 \ell$ (the locally integrated dissipation rate would then be $\epsilon_{2\ell}$). In a spherical integration over ${\bf r}$ of powers of the velocity difference $[u_i({\bf x}+{\bf r})-u_i({\bf x})]$, only the first term is affected by filtering or averaging over the spherical shell, while the center velocity $u_i({\bf x})$ remains fully local. Note that in the GKHE formalism, the averaging affect both end-point velocities in the same way, and both become averaged at scale $\ell$ in a formally symmetric way. 

An early connection between structure functions and filtering approaches was developed by   \cite{vreman1994realizability}. In the Vreman analysis, the structure function is defined based on the difference of velocity $u_i({\bf x}+{\bf r})$ and the locally filtered velocity $\tilde{u}_i$ centered at ${\bf x}$. Spherical integration of $(u_i({\bf x}+{\bf r}) - \tilde{u}_i)^2$ over a sphere of radius $\ell/2$ then yields equivalence with the SGS kinetic energy at scale $\ell$. But 
$(u_i({\bf x}+{\bf r}) - \tilde{u}_i)^2$ does not equal the usual structure function definition, now due to a mixture of filtered and unfiltered quantities at two points even before local filtering. 

Another interesting approach was presented in \cite{constantin94onsager} and connected to the LES filtering approach by \cite{eyink1995local}, \cite{eyink2006multi} (equations 2.12-2.14). 
In fact as recounted in the review by \cite{eyink2006onsager}, early unpublished work by Onsager anticipated such expressions half a century prior. Written in terms of the sharp spherical filter we use here, the  expression for the trace of the SGS stress reads
\begin{equation}
  \tau_{ii}({\bf x}) = \frac{1}{V_\ell} \iiint_{V_\ell} 
[u_i({\bf x}+{\bf r})-u_i({\bf x})]^2 \, d^3{\bf r}
-\left(\frac{1}{V_\ell}\iiint_{V_\ell} 
[u_i({\bf x}+{\bf r})-u_i({\bf x})] \, d^3{\bf r} \right)^2.
\label{eyinktauii}
\end{equation}
This equation represents an exact relationship between two-point structure functions and the subgrid-scale kinetic energy. But for the RHS to correspond to structure functions up to scale $\ell$, the integration must be done over a sphere of radius $\ell$ and thus a filtering scale of $2\ell$ for the stress tensor in the filtering formulation.  The suggested relationship then appears to be between subgrid-scale stress kinetic energy at scale $2\ell$ and structure functions up to two-point separations $\ell$ but averaged over a local domain of size $2\ell$, similarly as in the \cite{duchon2000inertial} approach. Note that while each of the terms in Eq. \ref{eyinktauii} is also a mixture of filtered and unfiltered velocities, the subtraction cancels the  local term and restores the fully filtered property inherent in the definition of $\tau_{ii}$. 

While not expecting qualitatively different results (except perhaps using the diameter instead of the radius as a name for ``scale''), we here continue our focus on the more ``symmetric'' formulation by Hill, with fixed position ${\bf x}$ specified at the midpoint between two points separated by vector ${\bf r}$ whose magnitude then spans up to scale $\ell$ (or integration radius ${\bf r}_s$ up to radius $\ell/2$).  

\section{Comparisons between kinetic energies and cascade rates/fluxes}
\label{sec:compstats}
 
In this section, we provide comparisons of local kinetic energies in the structure function formalism, $k_{{\rm sf},\ell}$, with that in the filtering formalism,
$k_{{\rm sgs},\ell}$. We also compare the local energy cascade rates $\Phi_\ell$ and $\Pi_\ell$. We consider data from a direct numerical simulation (DNS) of forced isotropic turbulence at $R_\lambda = 1{,}250$ (the Taylor-scale Reynolds number) that used 8,192$^3$ grid points \citep{yeung2012dissipation}. 
The analysis is performed at three length-scales in the inertial range, $\ell=30\eta, \,45\eta, \,60\eta$ 
where $\eta=(\nu^3/ \langle \epsilon \rangle)^{1/4}$ is the (average) Komogorov scale.   
Quantities are evaluated directly from their definitions using spherical integration. 
To compute volume spherically filtered quantities such as $\epsilon_\ell$, $k_{{\rm sgs},\ell}$, $k_{{\rm sf},\ell}$ or $\tau_{ij}$ from data, we fix the middle point coordinate ${\bf x}$ in the physical domain. Subsequently, we download data in a cubic domain using the JHTDB's cutout service in a cube of size equal to $\ell$. The data are then multiplied by a spherical mask (filter) to evaluate local filtered quantities.  The velocity components are obtained by utilizing pre-computed Getfunctions from the Johns Hopkins Turbulence database (JHTDB) including spatial interpolation, as explained in more detail in Appendix \ref{appA}. For surface averages, we discretize the outer surface of diameter $\ell$ into 500 points  pairs ($+$ and $-$ points) that are approximately uniformly distributed on the sphere.  The accuracy of this method of integration has been tested by increasing the number of points used in the discretization. Data on the specified points are obtained from the database using 8th-order Lagrange spatial interpolation.

\begin{figure}
 \centering
  \includegraphics[scale=0.35]{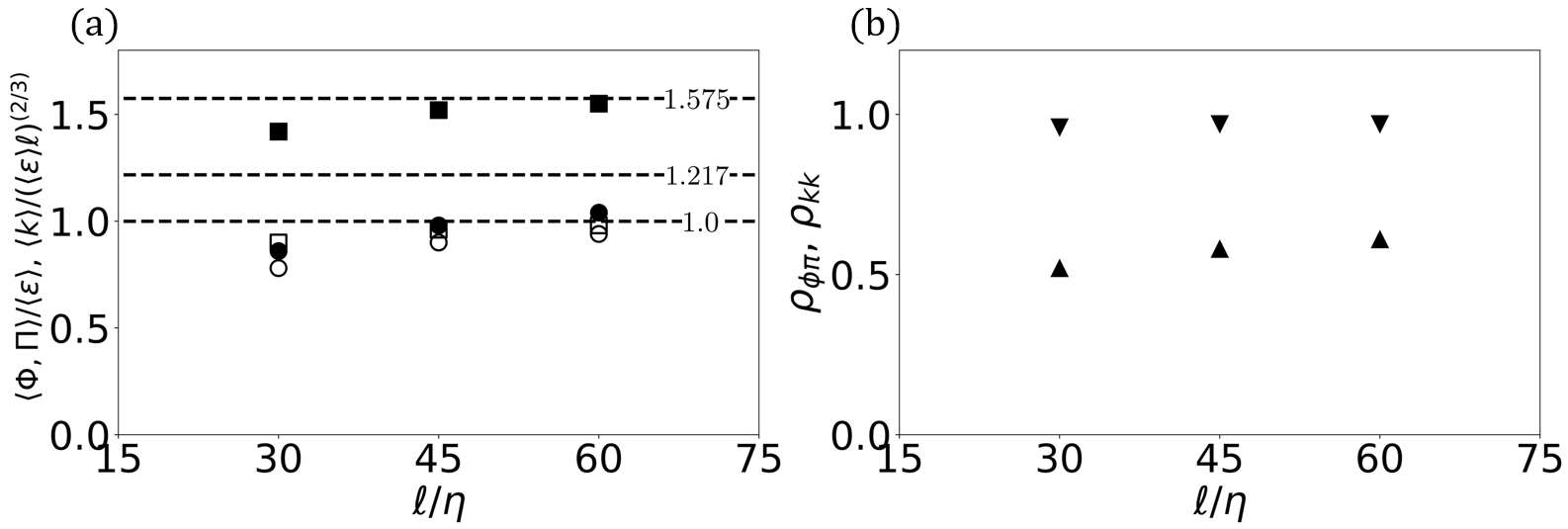}
    \caption{
    Panel (a) shows normalized mean kinetic energies and mean cascade rates as function of three filter scales for the $R_\lambda = 1250$ DNS isotropic turbulence dataset. Specifically, closed squares show $\langle k_{{\rm sf},\ell} \rangle / (\langle \epsilon \rangle \ell)^{2/3}$ while closed circles show $\langle k_{{\rm sgs},\ell} \rangle / (\langle \epsilon \rangle \ell)^{2/3}$.  Open squares show $\langle \Phi_\ell  \rangle / \langle \epsilon \rangle$ 
    while open circles show $\langle \Pi_\ell  \rangle / \langle \epsilon \rangle$. The horizontal lines show the expected asymptotic values in the inertial range for mean kinetic energies in the structure function formulation (1.575) and in the filtering formulation (1.217), while the expected energy cascade rates equal unity.  Panel (b) shows the correlation coefficients between kinetic energies ($\rho_{kk}$, downward triangles) and between cascade rates ($\rho_{\Phi \Pi}$,upward triangles).}
    \label{meankkpp}
\end{figure}

Overall mean values are obtained at the three scales and are plotted in Fig. \ref{meankkpp} (a).  The results for kinetic energy for the structure function approach are consistent with the analytical evaluation (see Appendix \ref{appB}). For the SGS kinetic energy, the numerical results fall below the theoretical inertial range prediction, due to effects of the viscous range that reduces the amount of SGS kinetic energy even at scales much larger than the Kolmogorov scale. The average values of the spectral fluxes agree very well with the averaged rate of viscous dissipation.

\begin{figure}
 \centering
  \includegraphics[scale=0.37]{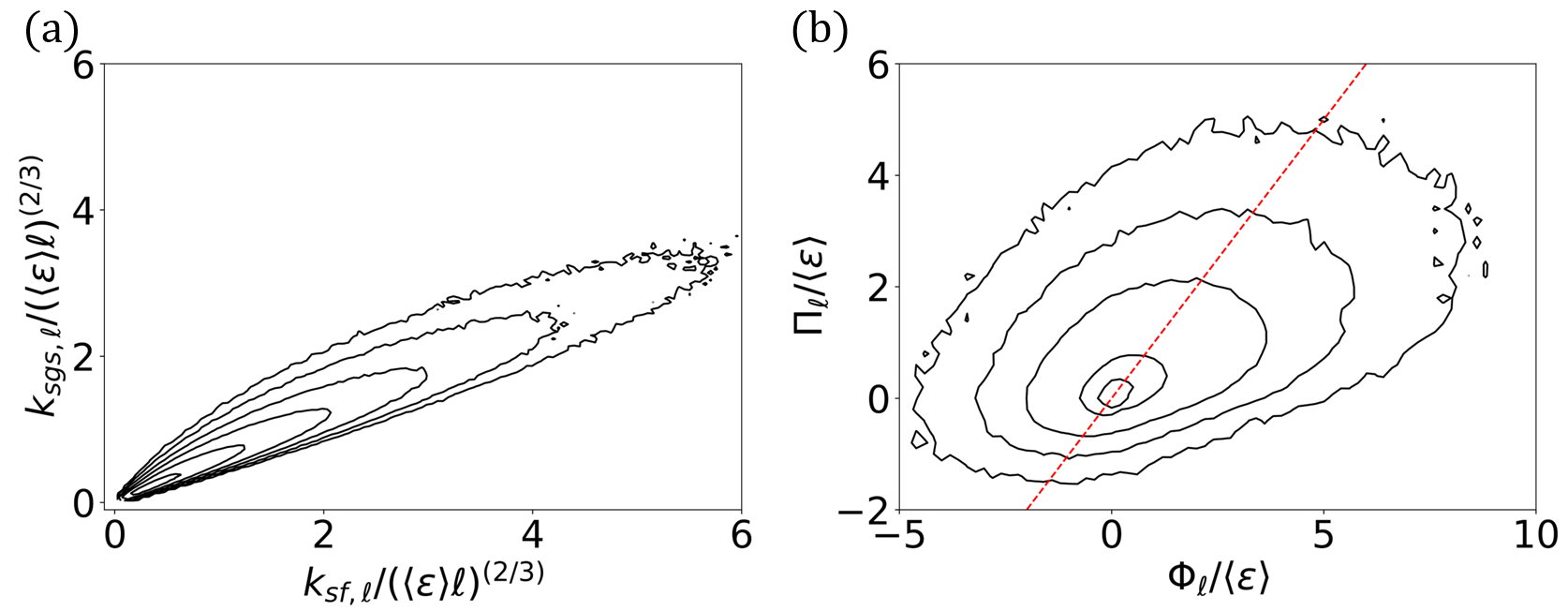}
    \caption{Joint PDFs of (a) $k_{{\rm sgs},\ell}$ and $k_{{\rm sf},\ell}$ with values $0.01, 0.03, 0.1, 0.3, 1, 3$ (b) $\Pi_\ell$ and $\Phi_\ell$ with values $0.001, 0.003,0.01, 0.3, 0.1, 0.3$, at scale $\ell = 45 \eta$ measured in DNS of isotropic turbulence at $R_\lambda$=1250. The red dash line represents a $45$ degree slop line. 
    The data and the editable notebook can be found at:  \href{https://cocalc.com/share/public_paths/27d603175bdd529d2ae4d091a3b56eb2f88eecc1}{https://cocalc.com/.../Figure3}.
    }
    \label{jpdf}
\end{figure}

Figure \ref{jpdf}(a) shows the joint PDF of $k_{{\rm sf},\ell}$ and $k_{{\rm sgs},\ell}$ at scale $\ell = 45 \eta$. The correlation coefficient between both quantities is $\rho_{kk} = 0.97$ (Fig. \ref{meankkpp} (b)).   Similarly, Figure \ref{jpdf}(b) shows the joint PDF of $\Pi_{\ell}$ and $\Phi_{\ell}$, also at scale $\ell = 45 \eta$ for the same dataset.  The correlation coefficient between both quantities is measured to be $\rho_{\Phi \Pi}=0.58$ (Fig. \ref{meankkpp} (b)), significantly lower than for the energies but still appreciable. It can be seen that negative values occur for both $\Pi_{\ell}$ and $\Phi_{\ell}$, although  it appears that $\Phi_{\ell}$ has more variability and larger negative excursions than $\Pi_{\ell}$. As summarized in \S \ref{sec:intro}, the relevance of locally negative values of $\Pi_{\ell}$ to the flow physics remains unclear, especially given the fact that upon averaging, the quantity becomes positive. Conversely, the quantity $\Phi_{\ell}$ has a clearer local interpretation, in the sense that locally negative values can clearly be interpreted as kinetic energy (local $\delta u_i^2/2)$) showing a net flux into a sphere of diameter $\ell$ in scale space, i.e. becoming associated with energy at smaller scales. Its overall average also is positive. An interesting question is whether negative values of $\Pi_{\ell}$ or $\Phi_{\ell}$ survive under some type of statistical averaging. In the next sections we use conditional averaging to quantify the importance of negative values (inverse local cascade, or backscatter). 

\section{Conditional averaging based on local dissipation}
\label{sec:compKRSH}

In this section, we compare conditionally averaged cascade rates/fluxes for both the structure function and filtering formulations, conditioned on $\epsilon_\ell$, i.e, $\langle \Phi_\ell | \epsilon_\ell \rangle $ and  $\langle \Pi_\ell | \epsilon_\ell \rangle $. 
According to KRSH \citep{kolmogorov1962refinement}, the statistical properties of velocity increments depend on the local average dissipation within a sphere of scale $\ell$, rather than being determined by the globally averaged dissipation. Written in terms of the quantities of present interest, the KRSH would read
$\langle \Phi_\ell | \epsilon_\ell \rangle = \epsilon_\ell$ since $\Phi_\ell$ is fully determined by the velocity increments envisioned in the KRSH. Loosely extending the KRSH arguments to the filtering formalism would suggest    $\langle \Pi_\ell | \epsilon_\ell \rangle = \epsilon_\ell$.

\begin{figure}
 \centering
    \includegraphics[scale=0.35]{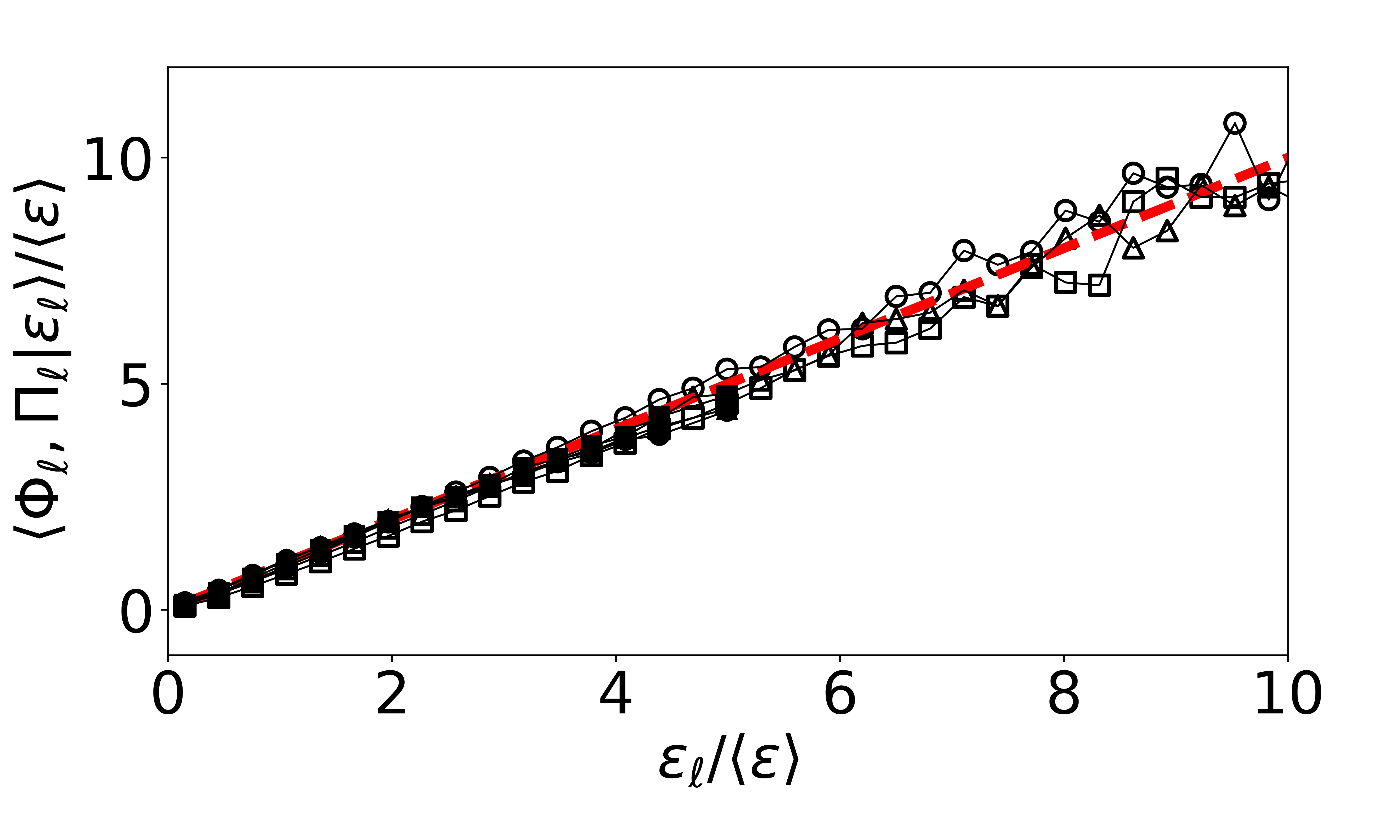}  
    \caption{Conditional averages of $\Phi_\ell$ and $\Pi_\ell$ based on local dissipation $\epsilon_\ell$, i.e., $\langle \Phi_\ell |\epsilon_\ell\rangle$ (black symbols and lines), $\langle \Pi_\ell |\epsilon_\ell\rangle$ (open symbols and lines). The red line indicates the value of $\epsilon_\ell$.  Different symbols denote different scales  $\ell/\eta = 30 $ (squares), $45$ (triangles) and $60$ (circles). All values are normalized with the globally averaged rate of dissipation $\langle \epsilon\rangle$.}
    \label{Fr_epr_304560k}
\end{figure}

In order to assess this hypothesis, we evaluate the conditional averages based on the same dataset described before. Results for  $\langle \Phi_\ell |\epsilon_\ell\rangle$ and $\langle \Pi_\ell |\epsilon_\ell\rangle$ are shown in Figure \ref{Fr_epr_304560k}. Results for the three scales considered are included. As can be seen, the plot shows a close agreement between both $\langle \Phi_\ell |\epsilon_\ell \rangle$ and $\langle \Pi_\ell |\epsilon_\ell\rangle$, with $\epsilon_\ell$. It is important to note that $\Phi_\ell$ and $\Pi_\ell$ are conditioned on the exact same values of $\epsilon_\ell$. The similarities and differences observed in Figure \ref{Fr_epr_304560k} indicate that $\Phi_\ell$ and $\Pi_\ell$ share many properties (same conditional averages)  but they are not identical. For instance, it is clear from Fig. \ref{jpdf} (b)  that the variance of $\Phi_\ell$ exceeds that of $\Pi_\ell$, even though their mean values are the same. 
 
In general, the behaviors of both $\langle \Phi_\ell | \epsilon_\ell \rangle$ and $\langle \Pi_\ell | \epsilon_\ell \rangle$ confirm the validity of the KRSH in the present context. More detailed analysis of the KRSH for $\Phi_\ell$ and connections to Eq. \ref{ins_GKHE_noint} are reported in \cite{yao2023kolmogorov}

\section{Conditional averaging based on large-scale velocity gradients}
\label{sec:compvelgrad}

In this section, we explore correlations between 
properties of the velocity gradient tensor filtered at scale $\ell$ and the two definitions of energy cascade rate/flux. The velocity gradient tensor encapsulates information about fluid deformation and rotation and connections to the energy cascade have been studied extensively. Already   \cite{bardina1985effect} examined the impact of rotation on homogeneous isotropic turbulence (HIT) and observed that rotation decreases the dissipation (cascade) rate while increasing the length scales, suggestive of  inverse energy cascade effects.   \cite{goto2008physical} investigated  physical mechanisms underlying forward energy cascade and argued that forward cascade can be triggered in regions characterized by strong strain between two large-scale tubular vortices. 
The role of the filtered gradient tensor for energy cascade was first explored numerically in \cite{borue1998local} and experimentally in \cite{van2002effects} building on the ``Clark model'' that approximates features of the subgrid-scale tensor using Taylor-series expansion.  
Recent studies by  \cite{johnson2020energy, johnson2021role} and \cite{carbone2020vortex} have significantly expanded on such analyses and examined the roles of self-strain amplification and vortex stretching driving the forward energy cascade process.  For inverse cascade, a vortex thinning mechanism may be at play \citep{johnson2021role}. 

A first level of characterization of the properties of the velocity gradient tensor are its invariants. To characterize deformation and rotation, we evaluate the strain and rotation invariants from data, defined according to
\begin{equation}
    S_\ell^2({\bf x},t) = \tilde{S}_{ij}\tilde{S}_{ij}, ~~~~~~
    \Omega_\ell^2({\bf x},t) = \tilde{\Omega}_{ij}\tilde{\Omega}_{ij},
\end{equation}
where $S_{ij}$ and $\Omega_{ij}$ are the symmetric and antisymmetric parts of the velocity gradient tensor $A_{ij}=\partial u_i/\partial x_j$ and the tilde denotes, as before, spherical tophat filtering over a ball of diameter $\ell$.
For consistency with prior literature, these values will be normalized by the overall average $\langle Q_w\rangle = \frac{1}{2}\langle\Omega_\ell^2\rangle$ (equal to 
  $\frac{1}{2}\langle S_\ell^2\rangle$ in isotropic turbulence).

A more detailed characterization of the statistics of velocity gradients involves the invariants $Q$ and $R$ \citep{vieillefosse1982local}. It is well-known that the joint probability density function (JPDF) of $Q$-$R$ exhibits a characteristic tear-drop shape \citep{chong1990general,meneveau2011lagrangian}, from which flow topology information such as vortex stretching and compression can be inferred \citep{chong1990general, borue1998local, luthi2009expanding,danish2018multiscale}. 
These two invariants (at scale $\ell$) are defined as usual according to
\begin{equation}
    Q_\ell({\bf x},t) = -\frac{1}{2} \tilde{A}_{ij}\tilde{A}_{ji}, ~~~~~~
    R_\ell({\bf x},t) = -\frac{1}{3} \tilde{A}_{ij}\tilde{A}_{jk}\tilde{A}_{ki}.
\end{equation}
Under the assumption of restricted Euler dynamics \citep{meneveau2011lagrangian}, the transport equation for the velocity gradient tensor leads to ${d Q_\ell}/{d t} = -3R_\ell$ and ${d R_\ell}/{d t} = -\frac{2}{3}Q_\ell^2$ \citep{cantwell1992exact}. The quantity $R_\ell$ can thus be considered as the  (negative) rate of change of $Q_\ell$ and contains both vortex stretching and strain self-stretching mechanisms \citep{johnson2021role}.  In our comparative investigation of energy cascade rates, conditional averaging based on the four invariant quantities $S^2_\ell$, $\Omega^2_\ell$, $Q_\ell$ and $R_\ell$ will be undertaken.

We begin with qualitative visualizations of the fields in small subsets of the domains analzyed.
Panel (a) and (b) of Figure \ref{F_PI_O_S_45} depict a sample instantaneous field of $\Phi_\ell$ and $\Pi_\ell$ respectively, highlighting regions of strong local forward cascade (indicated by solid red circles) and strong inverse cascade (indicated by dashed circles). The correlation between these two variables is evident, the computed correlation coefficient between the snapshots is 0.64. On both panels (a) and (b) the fluxes are normalized by the global averaged dissipation $\langle \epsilon \rangle$. As already noted based on the joint PDFs, there are differences between $\Phi_\ell$ and $\Pi_\ell$. The maximum magnitude of the positive cascade rate in $\Phi_\ell$ is about twice that of $\Pi_\ell$, while the magnitude of the negative cascade rate in $\Phi_\ell$ is about 3 to 4 times larger. Since $\langle \Phi_\ell \rangle \sim \langle \Pi_\ell \rangle \sim \langle \epsilon_\ell \rangle \sim \langle \epsilon \rangle$, the significant different maximum values indicates $\Phi_\ell$ is more variable and intermittent than $ \Pi_\ell$. Also, $\Phi_\ell$ exhibits somewhat finer-scale spatial features. 
  
\subsection{Conditional statistics based on strain rate $(S^2_\ell)$ and rotation rate $(\Omega^2_\ell)$}

 \begin{figure}
 \centering
  \includegraphics[scale=0.6]{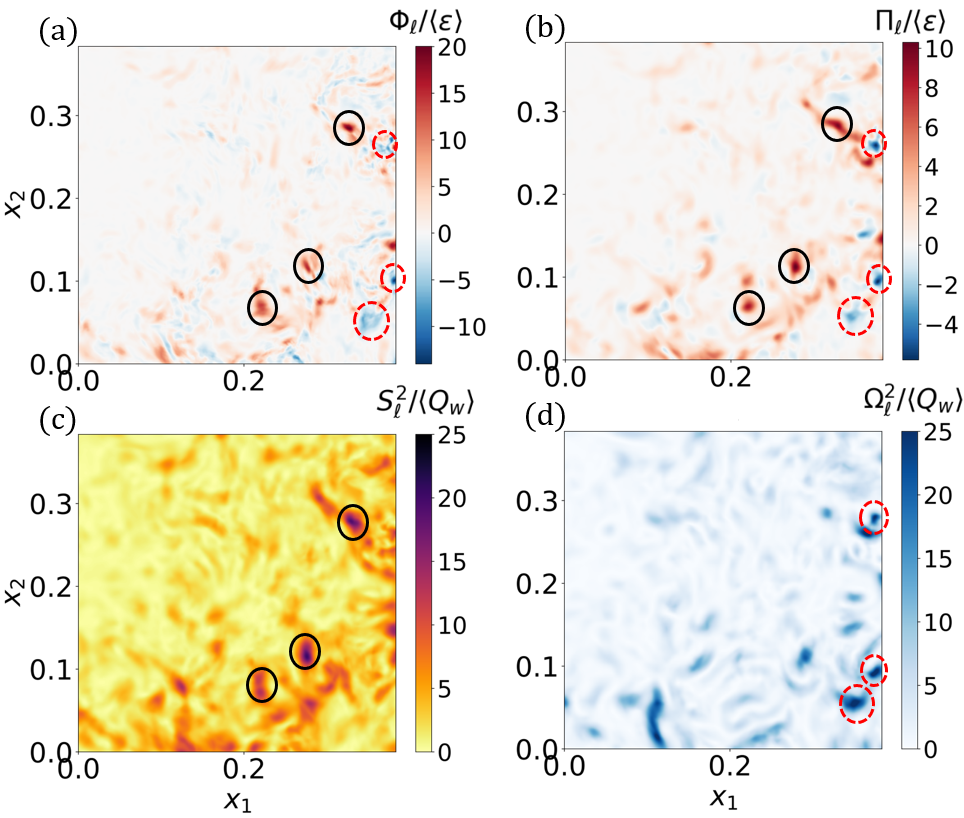}
    \caption{Panels (a), (b), (c), (d) are instantaneous $\Phi_\ell$, $\Pi_\ell$, $S_\ell^2$ and $\Omega_\ell^2$ field with $\ell=45\eta$ in a $750\eta \times 750 \eta$ domain ($500\times500$ points of the DNS grid. The black solid circles in (a) and (b) are located at  strong local forward cascade region, which is correlated to the strong local strain rate marked in the black circle in panel (c). The red dashed circles in (a) and (b) are located at strong local inverse cascade regions (negative energy fluxes), which appear qualitatively correlated to relatively strong local rotation rates marked in the red dashed circles in panel (d).}
    \label{F_PI_O_S_45}
\end{figure}

Panels (a) and (b) in figure \ref{F_PI_O_S_45} show distinct regions including both local forward (red area) and inverse (blue area) cascade rates. It is visually apparent that the presence of a strong local forward energy cascade is associated with increased local strain rate, as indicated by the solid red circle in both panels (a) and (b) of Figure \ref{F_PI_O_S_45} and the corresponding black solid circle in panel (c). 
Similarly, a strong local inverse energy cascade is observed alongside a significant local rotation rate, depicted by the dashed red circle in panels (a) and (b) of Figure \ref{F_PI_O_S_45} and the corresponding red dashed circle in panel (d). The strong correlation between forward cascade and local straining is consistent with multiple earlier observations and prior works in the literature (e.g. \cite{borue1998local} and recently \cite{johnson2021role, carbone2020vortex}).  We focus attention on the regions with negative energy cascade rates.  Conditional averaging can elucidate the statistical significance of these regions. Specifically, we inquire whether there are  large-scale flow local features as characterized by the filtered velocity gradient invariants that are systematically accompanied by inverse cascade, i.e., negative $\Phi_\ell$.  Thus we perform conditional averaging of $\Phi_\ell$ based on the invariants $S_\ell^2$ and $\Omega_\ell^2$ and repeat the analysis for the SGS energy flux quantity $\Pi_\ell$. 

 \begin{figure}
 \centering
  \includegraphics[scale=0.45]{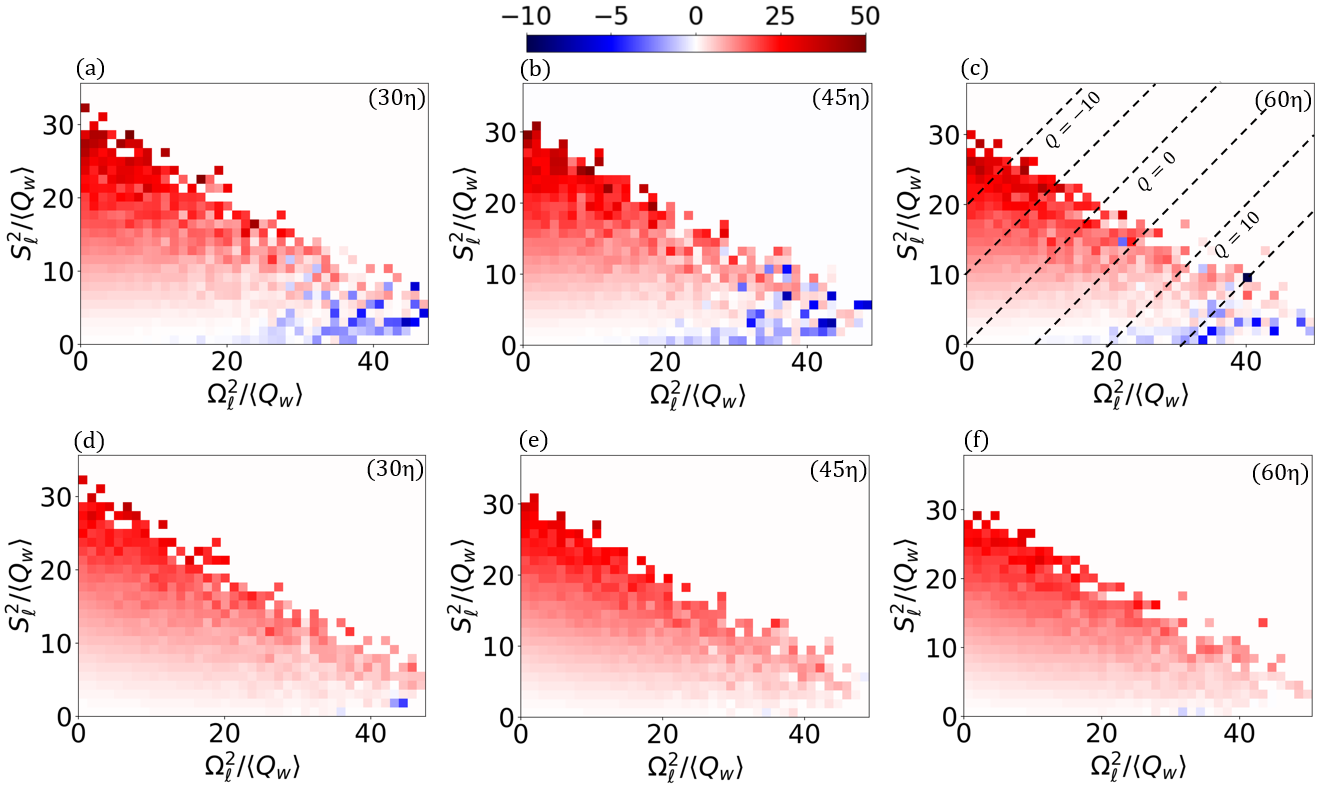}
    \caption{ Panels (a), (b), (c) are $\langle \Phi_\ell | S^2_\ell, \Omega^2_\ell\rangle$; panels (d), (e), (f) are $\langle \Pi_\ell | S^2_\ell, \Omega^2_\ell\rangle$ at $\ell = 30\eta, 45\eta, 60\eta$. The black dashed line in panel (c) represent iso-line of  $Q_\ell \equiv -\frac{1}{2}(S_\ell^2 - \Omega_\ell^2)$. $\Phi_\ell$ and $Pi_\ell$ are normalized by $\langle \epsilon \rangle$; $S^2_\ell$ and $\Omega^2_\ell$ are normalized by $\langle Q_w \rangle = \frac{1}{2}\langle \Omega_\ell^2 \rangle$.}
    \label{Fr_S2O2_304560k}
\end{figure}

Figure \ref{Fr_S2O2_304560k} shows the joint conditionally-averaged $\Phi_\ell$ and $\Pi_\ell$ based on $S^2_\ell$ and $\Omega^2_\ell$, denoted as $\langle \Phi_\ell | S^2_\ell, \Omega^2_\ell\rangle$ and $\langle \Pi_\ell | S^2_\ell, \Omega^2_\ell\rangle$, respectively. The analysis is performed by computing averages over two million randomly distributed points ${\bf x}$. In the presented results, $\Phi_\ell$ is normalized by $ \langle \epsilon \rangle$, while $S^2_\ell$ and $\Omega^2_\ell$ are normalized by $\langle Q_w \rangle = \frac{1}{2}\langle \Omega^2_\ell \rangle$. Panels (a), (b), and (c) of figure \ref{Fr_S2O2_304560k} present the joint conditionally-averaged $\langle \Phi_\ell | S^2_\ell, \Omega^2_\ell\rangle$ at three different length scales, namely $r = 30 \eta, 45\eta, 60 \eta$. Panels (a), (b), and (c) highlight the dominance of the forward cascade by the extensive red region. This magnitude is manytimes larger than the maximum magnitude observed in the blue region, representing the inverse cascade. The red region covers a wide range of $S^2_\ell$ and $\Omega^2_\ell$ values, consistent with the expectation that the global average would favor a forward cascade ($\langle \Phi_\ell>0 \rangle$). The highest positive values of $\langle \Phi_\ell | S^2_\ell, \Omega^2_\ell\rangle$ correspond to high strain rates and low rotation rates, and they decrease as the strain rate decreases. Interestingly, the inverse cascade appears explicitly in the lower-right corner of the plots, where the rotation rate is strong but the strain rate is weak. It is worth noting that the conditionally averaged values shown in Figure \ref{Fr_S2O2_304560k} reflect the combined outcome of the forward and inverse cascades. Consequently, in specific regions characterized by distinct strain and rotation rates, events with forward and inverse cascades can  cancel each other out. Only in the lower right corner is there an indication of net inverse cascade when the cascade rate is defined using the structure function approach. 

In panel (c), we superimpose  dashed lines representing the isolines of  $Q_\ell$, with the $Q_\ell = 0$ line indicating the condition of equal strain and rotation rates. The parallel dashed lines correspond to $Q_\ell=-10$, $-5$, $0$, $5$, $10$, and $15$, respectively. The $Q_\ell=15$ line appears near the boundary separating the red and blue regions. However, the boundary of the blue region   does not appear to align well with the $Q_\ell$  isoline. This observation suggests that $Q_\ell$ might be not enough to provide an adequate threshold for distinguishing the net forward and inverse cascade regions.

Panels (d), (e), and (f) in Figure \ref{Fr_S2O2_304560k} present results for the joint conditionally-averaged $\Pi_\ell$ based on $S^2_\ell$ and $\Omega^2_\ell$, corresponding to the same filter scales as panels (a), (b), and (c). It is evident that  trends for the positive cascade rate (red region) for $\Pi_\ell$ closely resemble those  of $\Phi_\ell$, with the peak of the forward cascade occurring at a high strain rate and low rotation rate. The magnitude of the maximum forward cascade rate for $\Pi_\ell$ is slightly weaker compared to that of $\Phi_\ell$. The most significant difference is that only a few instances of blue squares are observed in regions characterized by strong rotation and weak strain, indicating that the overall predominance of the forward cascade persists regardless of the local values of $S^2_\ell$ and $\Omega^2_\ell$. These results  highlight some important statistical differences between $\Phi_\ell$ and $\Pi_\ell$.

 \begin{figure}
 \centering
  \includegraphics[scale=0.45]{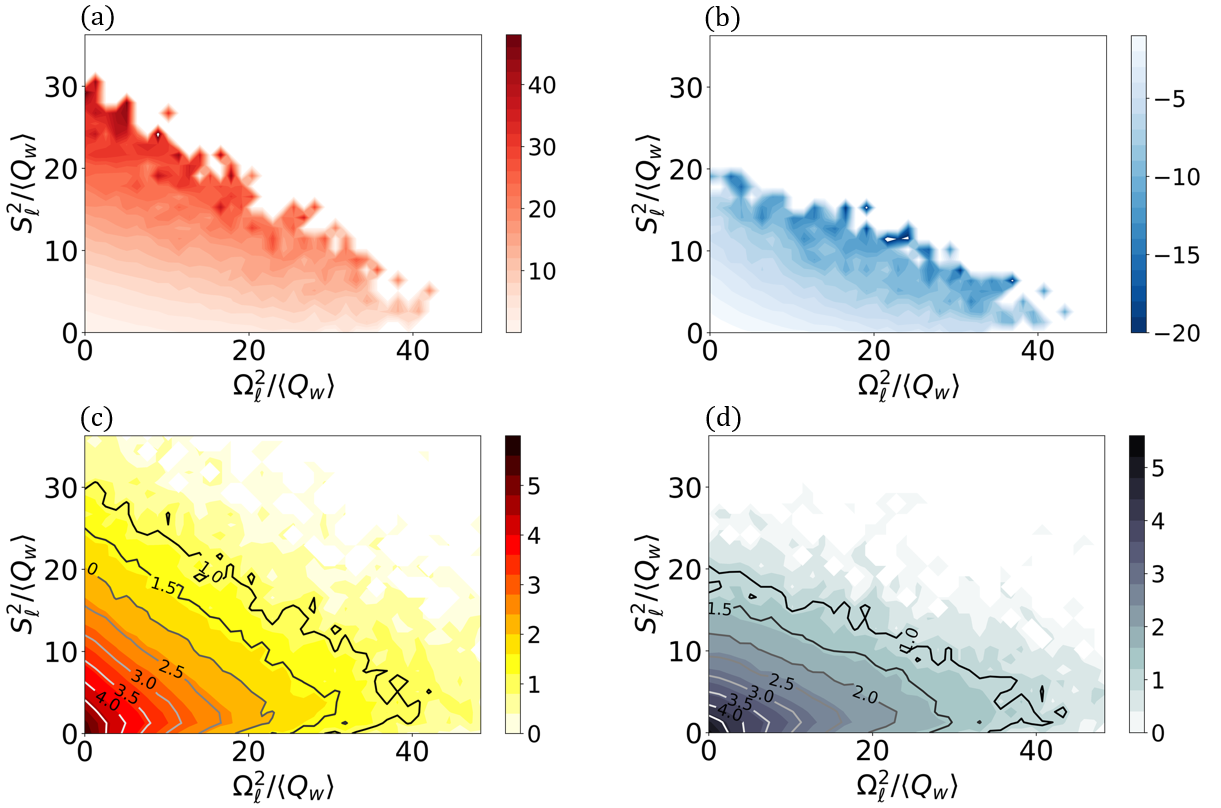}
    \caption{
    Panels (a) and (b) present the conditional averages of cascade rates obtained by sampling positive and negative signs, i.e, $\langle \Phi_\ell | S^2_\ell, \Omega^2_\ell, \Phi_\ell > 0 \rangle$ and $\langle \Phi_\ell | S^2_\ell, \Omega^2_\ell, \Phi_\ell < 0 \rangle$, respectively. Panel (c) and (d) display the logarithm base 10 of the number of samples on the $S^2_\ell, \Omega^2_\ell$ map (out of a total of $2\times 10^6$ samples). The isolines in panels (c) and (d) are the values corresponding to the contour.}
    \label{InandF_Fr_S2O2_45k}
\end{figure}

To develop a more detailed understanding of the inverse cascade region within $\langle \Phi_\ell | S^2_\ell, \Omega^2_\ell \rangle$, we performe a further analysis by dividing the samples based on $\Phi_\ell > 0$ and $\Phi_\ell < 0$ for $r = 45 \eta$ (results for $r = \{30 , 60\} \eta$, not shown, are similar). We then  calculate the conditional average of these separated samples considering $S^2_\ell$ and $\Omega^2_\ell$, denoted as $\langle \Phi_\ell | S^2_\ell, \Omega^2_\ell, \Phi_\ell > 0 \rangle$ and $\langle \Phi_\ell | S^2_\ell, \Omega^2_\ell, \Phi_\ell < 0 \rangle$. The results are presented in Figure \ref{InandF_Fr_S2O2_45k}. From panel (a), it can be observed that the forward cascade clearly increases with $S^2_\ell$, with the highest values of $\Phi_\ell$ concentrated in the upper-left corner. It increases also with $\Omega^2_\ell$ but less rapidly. Combined, the trend seems to be an increase  
roughly proportional to $ \sim S^2_\ell + 0.75 \, \Omega^2_\ell$.
Differently, panel (b) illustrates that the inverse cascade is roughly proportional to $ \sim S^2_\ell + 0.5 \, \Omega^2_\ell$, i.e. shallower isolines extending more in the horizontal direction than in the vertical compared to the forward cascade case shown in (a). This observation elucidates why the strongest red region in panel (b) of Figure \ref{Fr_S2O2_304560k} emerges at the largest $S^2_\ell$, while below this threshold, the forward cascade events progressively weaken and are gradually canceled out by the inverse cascade. Finally, in regions characterized by a weak strain rate and strong rotation rate, the inverse cascade becomes the dominant contribution.

Panel (c) and (d) display the distribution of the number of samples corresponding to positive and negative cascade rates in logarithmic scale (out of the 2 million samples (balls) considered). Our focus is specifically directed towards the bottom-right corner of the plots, which corresponds to the region where the inverse cascade is observed in panel (b) of Figure \ref{Fr_S2O2_304560k}. Interestingly, we observe that at $\Omega^2_\ell/\langle Q_w \rangle \approx 40$ and $S^2_\ell/\langle Q_w \rangle < 10$, the number of samples representing both inverse and forward cascade rates is roughly equivalent, falling within the range of $10^{1}$ to $10^{1.5}$. This implies that within this region, the magnitude of the inverse cascade must be significant to achieve net negative values for the conditional average.  Still, for the  conditions with net inverse cascade, the number of occurrences for both forward and inverse cascade rates is quite small, on the order of only $1/10^5$ of the total samples, indicating a very low frequency. This observation suggests that the inverse cascade region depicted in Figure \ref{Fr_S2O2_304560k} must be primarily attributed to very rare but intense events.

\subsection{Conditional statistics based on $Q_\ell$ and $R_\ell$ invariants}

In the context of the $\Omega^2_\ell$ and $S^2_2$ map shown in Figure \ref{Fr_S2O2_304560k}, we observe the presence of a distinct inverse energy cascade in the region characterized by strong rotation but weak strain (corresponding to large $Q$ values) for $\Phi_\ell$. However, such observations did not hold for $\Pi_\ell$. But these results  do not preclude the possibility that  net forward and inverse cascade may be associated with other invariants of the filtered velocity gradient tensor $\tilde{A}_{ij}$.

\begin{figure}
 \centering
  \includegraphics[scale=0.45]{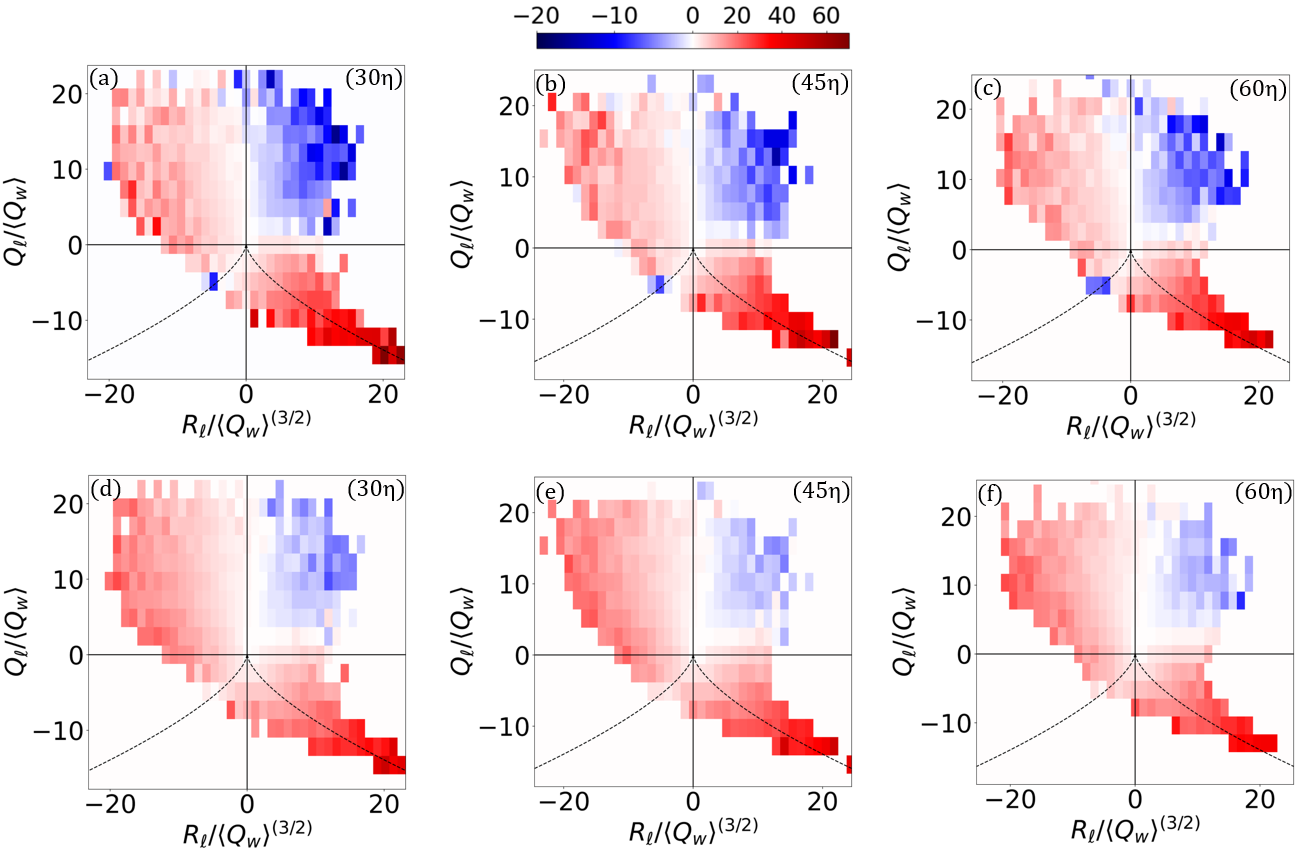}
    \caption{ Panels (a), (b), (c) are $\langle \Phi_\ell | Q_\ell, R_\ell\rangle$; panels (d), (e), (f) are $\langle \Pi_\ell | Q_\ell, R_\ell\rangle$ at $\ell = 30\eta, 45\eta, 60\eta$. The black dotted lines separate four quadrants and the two lines $Q=-(\frac{27}{4} R^2)^{1/3}$ (Viellefosse lines) are also shown. $\Phi_\ell$ and $\Pi_\ell$ are normalized by $\langle \epsilon \rangle$; $Q_\ell$ are normalized by $\langle Q_w \rangle$ and $R_\ell$ are normalized by $\langle Q_w \rangle^{(3/2)}$. The data and editable analysis code that generated these joint PDFs (for the case at $\ell=45\eta$) can be found at: \href{https://cocalc.com/share/public_paths/9622d6e480f8491f5ab253198042d36deecb36e6}{https://cocalc.com/.../Figure8}.}
    \label{Fr_QR_304560k}
\end{figure}

Figure \ref{Fr_QR_304560k} shows the joint conditional averaged $\langle \Phi_\ell | Q_\ell, R_\ell \rangle$ and $\langle \Pi_\ell | Q_\ell, R_\ell \rangle$ at three different scales, namely, $\ell = \{30, 45, 60\} \eta$. Across all figures, we can observe the distinctive teardrop shape pattern on the Q-R map, as reported in previous studies 
\citep{chong1990general,meneveau2011lagrangian}. The black solid lines are the boundaries, separating the four quadrants based on the sign of $Q$ and $R$. Notably, it becomes evident that both $\Phi_\ell$ and $\Pi_\ell$ exhibit a strong and dominant inverse cascade in the quadrant characterized by $Q>0$ and $R>0$, commonly referred to as the ``vortex compression" region. This observation is particularly interesting considering the absence of an observable inverse cascade for $\Pi_\ell$ in the $S^2_\ell$ and $\Omega^2_\ell$ map. Hence, these results indicate that the variables $Q$ and $R$ provide a more effective characterization to identify inverse cascade using conditional averaging. 

The region characterized by $Q<0$ and $R>0$, which corresponds to the strain-dominated region, exhibits the most pronounced forward cascade. This observation aligns with the findings 
of many prior analyses in the literature \cite{borue1998local,van2002effects,johnson2021role,carbone2020vortex} as well as those 
depicted in Figure \ref{Fr_QR_304560k}, further emphasizing that the strong local strain rate plays a crucial role in driving the forward energy cascade. We note that \cite{borue1998local,van2002effects} display conditional averages weighted by the joint PDF of $R$ and $Q$. In their results, there was hardly any indication of backscatter/inverse cascade in the  $Q>0$ and $R>0$ ``vortex compression" region, because the overall probability density of that region is smaller than the other regions. However, the unweighted conditional averaging represents the relevant values if the large-scale flow is in that particular state ($Q>0$ and $R>0$), and is therefore relevant to our analysis. 

\begin{figure}
 \centering
  \includegraphics[scale=0.55]{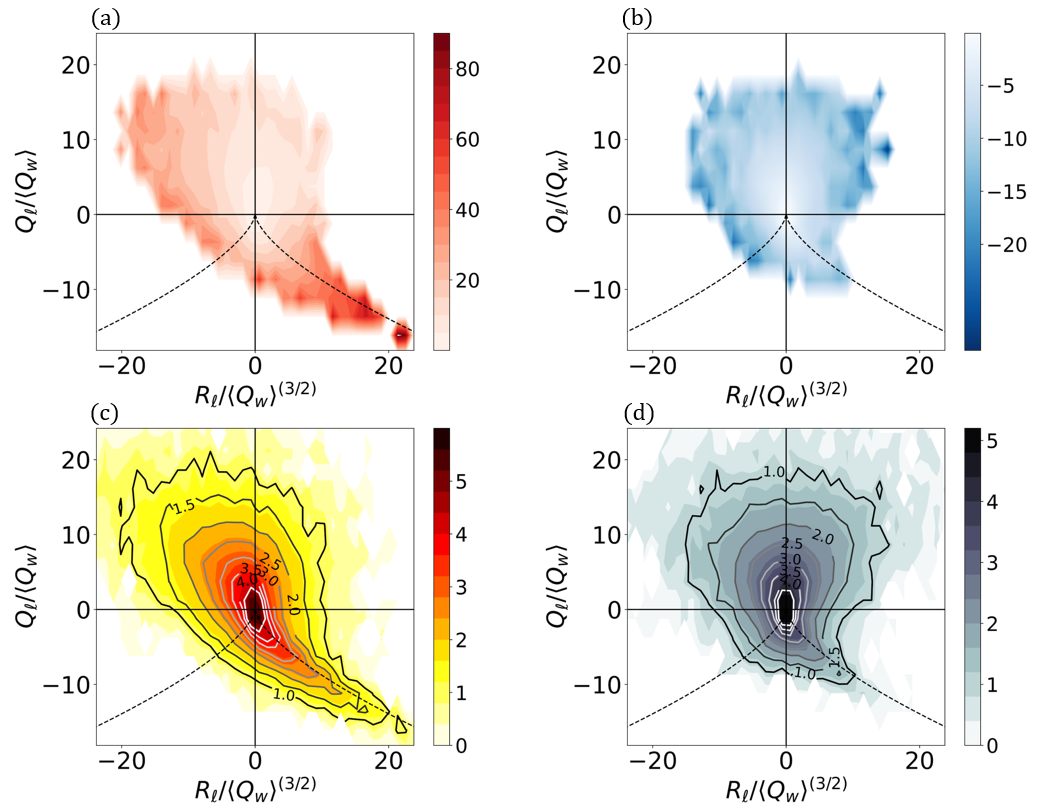}
    \caption{Panels (a) and (b) show $\langle \Phi_\ell | Q_\ell, R_\ell, \Phi_\ell > 0 \rangle$ and $\langle \Phi_\ell | Q_\ell, R_\ell, \Phi_\ell < 0 \rangle$. (c) and (d) display the logarithm base 10 of the number of samples on the $Q, R$ map. The isolines in panels (c) and (d) are the values corresponding to the contour.}
    \label{InandF_Fr_QR_45k}
\end{figure}

In a similar manner to Figure \ref{InandF_Fr_S2O2_45k}, we perform further conditional averaging also distinguishing positive and negative cascade rates. Figure \ref{InandF_Fr_QR_45k} (a) and (b) present $\langle \Phi_\ell | Q_\ell, R_\ell, \Phi_\ell > 0 \rangle$ and $\langle \Phi_\ell | Q_\ell, R_\ell, \Phi_\ell < 0 \rangle$ at $\ell = 45 \eta$. In the case of the inverse cascade, it is observed to occur in all four quadrants (panel (b)), with a more evenly distributed and symmetric presence in the upper two quadrants associated with $Q>0$, i.e, the rotation-dominated regions. The characteristic teardrop shape is less prominent and exhibits a shorter tail compared to the forward cascade (panel (a)). Regarding the forward cascade, it is evident that it is most dominant in the $Q<0, R>0$ quadrant, consistent with Figure \ref{Fr_QR_304560k}. However, in the $Q>0, R>0$ quadrant, the forward cascade is weaker and is overall canceled out by the stronger inverse cascade in that particular region.  

Panels (c) and (d) display the  distribution of number of samples of forward and inverse cascade rates, respectively. The shapes of the distributions align with Figure \ref{InandF_Fr_QR_45k} (a) and (b), but a majority of the samples are concentrated at the center, corresponding to small values of Q and R. This observation confirms that the strong instances of inverse cascade and forward cascade observed in Figure \ref{Fr_QR_304560k} are primarily determined by  infrequent but extreme events (intermittency).

\begin{figure}
 \centering
  \includegraphics[scale=0.5]{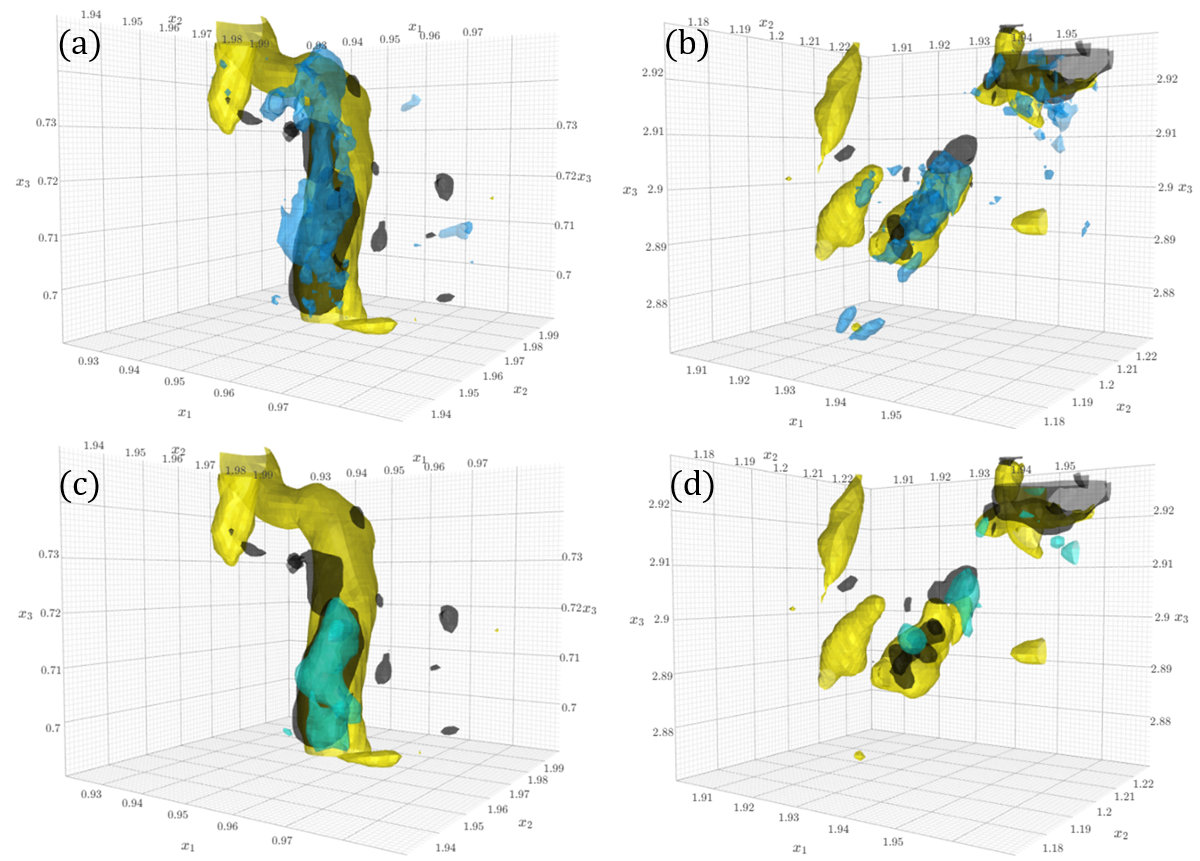}
    \caption{Panel (a) and (b) show
    isosurface of local $\Phi_\ell/\langle \epsilon \rangle = -60$ (light blue) and $Q_\ell/\langle Q_w \rangle = 20$ (yellow), and $R/\langle Q_w \rangle^{(3/2)} = 30$ (black) in two different 3D subdomains. Panel (c) and (d) show isosurface of local $\Pi_\ell/\langle \epsilon \rangle = -20$ (green-blue) in the same subdomians and isosurface of  $Q_\ell/\langle Q_w \rangle$ and $R/\langle Q_w \rangle^{(3/2)}$ as panel (a) and (b).  
    Interactive visualizations are available for each panel at: \href{https://cocalc.com/share/public_paths/7b9005ee7515fbe20bcdc224a3aca67b52799052/Figure10_a.html}{Panel(a)},  \href{https://cocalc.com/share/public_paths/7b9005ee7515fbe20bcdc224a3aca67b52799052/Figure10_b.html}{Panel(b)},   \href{https://cocalc.com/share/public_paths/7b9005ee7515fbe20bcdc224a3aca67b52799052/Figure10_c.html}{Panel(c)},  \href{https://cocalc.com/share/public_paths/7b9005ee7515fbe20bcdc224a3aca67b52799052/Figure10_d.html}{Panel(d)}.
    The link to the directory containing the visualization code and the 3D fields with these data can be found at: \href{https://cocalc.com/share/public_paths/7b9005ee7515fbe20bcdc224a3aca67b52799052}{https://cocalc.com/.../}.
    }
    \label{3Dviz_phi}
\end{figure}

Finally, to provide a visual impression of the spatial distribution of regions of negative $\Phi_\ell$ in the flow, in Fig. \ref{3Dviz_phi} we provide a 3D visualization of two instances in specific small subdomains of size $150^3$ gridpoints (i.e. $(225\eta)^3$ out of the overall $8192^3$ DNS domain. The selection of these subdomains was based on the condition that $Q_\ell/\langle Q_w \rangle > 15$ and $R_\ell/\langle Q_w \rangle^{(3/2)} > 15$ at the center of each subdomain such that the center is at a strong vortex compression region. We then calculate the values of $\Phi_\ell$, $\Pi_\ell$, $Q_\ell$ and $R_\ell$ at every second grid point. In panel (a) and (b) of Figure \ref{3Dviz_phi}, the light blue regions correspond to isosurface  of a large negative value of $\Phi_\ell/\langle \epsilon \rangle  = - 60$, indicating the presence of an inverse cascade with significant magnitude. Clearly, we can see that the occurrence of strong inverse cascade is closely associated with the presence of the vortices. Panel (a) depicts that large negative $\Phi_\ell$ appears near the center and not at the core of the yellow tube, although one should recall that $\Phi_\ell({\bf x},t)$ is defined locally as centered at ${\bf x}$ but represents the energy cascade into balls of diameter $45\eta$, i.e. comparable to the diameter of the vortex (yellow region) shown. The blue region is also largely connected with the black isosurface, ($R_\ell=20 
\langle Q_w\rangle^{3/2}$) indicating a strong ``vortex compression'' region  within the yellow tube. Interactive 3D versions of the figure that can be accessed following the links in the figure caption help elucidate the spatial structure. Panel (b) is an entirely different instance of similar conditions, showing a more broken up vortex and showing that $\Phi_\ell$ can also peak near the sides, and appear more scattered within the vortex. Coupled with the results shown in Fig. \ref{Fr_QR_304560k}, the visualizations suggest that the strong inverse cascade occurs along the large scale vortices, in regions of these vortices in which $R>0$, i.e. the vortex compression regions. We can also observe some yellow tubes, within which inverse cascade and compression are both absent. This is consistent to the statistics such that, when conditionally averaging in terms of $Q_\ell$ but irrespective of $R$, the inverse cascade becomes very week and almost non-existent. However, once one only considers $R>0$ regions, inverse cascade can be clearly observed in high vortical regions. 
 
We also show $\Pi_\ell /\langle \epsilon \rangle = -20$ (green-blueish  isosurface) in the corresponding 3D subdomains shown as panels (c) and (d) of Figure \ref{3Dviz_phi}. Clearly we can see that the green-blueish and black regions largely overlap within the yellow region in panel (c). In panel (d),  the overlapping between green-blueish, yellow and black region occurs at the center and right-top region of the subdomain, indicating strong negative $\Pi_\ell$ is also associated with strong vortex compression within high vortical region, consistent with Figure \ref{Fr_QR_304560k}. However, the patterns of the green SGS flux regions are smoother, consistent to the 2D visualisation in Figure \ref{F_PI_O_S_45}. 

Caution must be expressed that visualizations only provide qualitative impressions and more quantitative analysis requires structure-based conditional averaging such as recently undertaken in \cite{park2023coherent}. While such analysis is beyond the scope of the present paper, the conditional statistics presented in Fig. \ref{Fr_QR_304560k} already by themselves provide the strong statistically robust connection between cascade rate measures  and features of the large scale velocity gradient tensor. 
  
\section{Conclusions}
\label{sec:conclusions}

In this paper we explore, based on a DNS dataset of isotropic forced turbulence at a relatively high Reynolds number ($R_\lambda = 1250$), 
local features of the energy cascade. We compare two common definitions of the spatially local rate of kinetic energy cascade at some scale $\ell$. The first is based on the cubic velocity difference term appearing in the Generalized Kolmogorov-Hill equation (GKHE), in the structure function approach. The second is the subfilter-scale energy flux term in the transport equation for subgrid-scale kinetic energy, i.e., as used in filtering approach often invoked in LES. Particular attention is placed on interpretation and statistical robustness of observations of local negative structure-function energy flux and subfilter-scale energy flux. The notion and relevance of local inverse cascade or  ``backscatter'' has been open to debates in the literature. 
We argue that the interpretation of $\Phi_\ell({\bf x},t)$ as  a spatially local energy flux appears unambiguous because it arises naturally from a divergence term in scale space. And, the symmetric formulation of \cite{hill2001equations,hill2002exact} leads to the spherically averaged third-order structure function-based definition of a local cascade rate involving velocities at two points that are treated equally via angular averaging over the sphere.

The data confirm the presence of local instances where $\Phi_\ell({\bf x},t)$ is negative, i.e. indicative of local inverse cascade events in 3D turbulence. Flow visualizations show that spatially the inverse cascade events are often  located near the core of large-scale vortex structures. Comparable observations for the LES-based energy flux  
$\Pi_\ell({\bf x},t)$ (which also displays negative values at many locations in the flow as is well-known in the LES literature on ``backscatter'') show that $\Pi_\ell({\bf x},t)$ displays smoother and more blob-like features. Regarding the statistical significance of such observations, local observations from single realizations are extended using conditional averaging. Attention is placed first on relationships between the local cascade rate and the local filtered viscous dissipation rate $\epsilon_\ell({\bf x},t)$ that plays a central role in the classic KRSH \cite{kolmogorov1962refinement}. Results show that conditional averaging of both $\Phi_\ell({\bf x},t)$ and $\Pi_\ell({\bf x},t)$ eliminates negative values and that the conditional averages in fact equate $\epsilon_\ell({\bf x},t)$ to very good approximation, entirely consistent with KRSH predictions. 

The analysis then focuses on conditional averages of  $\Phi_\ell$ and $\Pi_\ell$ conditioned on properties of the filtered velocity gradient tensor properties, in particular four of its most invariants (strain and rotation rate square magnitude and the two $Q-R$ invariants). We find statistically robust evidence of inverse cascade as measured with $\Phi_\ell$  when both the large-scale rotation rate is strong and the large-scale strain rate is weak. When defined using $\Pi_\ell$, the conditional averaging based on large-scale strain and rotation rates  does not lead to any significant average backscatter.  When conditioning based on the $R$ and $Q$ invariants, significant net inverse cascading is observed for $\Phi_\ell$ in the ``vortex compression'' $R>0$, $Q>0$ quadrant. Qualitatively similar, but quantitatively much weaker trends are observed for the conditionally averaged subfilter scale energy flux $\Pi_\ell$. We recall that a multiscale decomposition of $\Pi_\ell$ in terms of velocity gradients at multiple scales \cite{johnson2020energy,johnson2021role} shows that $\Pi_\ell<0$ appears associated with a vortex-thinning mechanism occuring at smaller scales interacting with large-scale strains. 

In summary, present results show that locally negative values of kinetic energy fluxes at scale $\ell$ are observed for both the structure function and filtering approaches, and at least for the structure function approach, the interpretation  as a flux in scale space appears unambiguous. Regarding statistical robustness and potential net impact of such local observations, conditional averaging shows that such inverse cascade becomes the statistically dominant mechanism in regions in which  the turbulent motions at scales larger than $\ell$  are of the ``vortex compression'' ($R>0$ and $Q>0$) type. 

Future work should extend conditional averaging to more accurately reflect entire flow structures and their possible connections to local inverse cascade mechanisms. Other pointwise quantities such as helicity can also be explored.  It would also be instructive to connect present results with the multiscale decomposition of \cite{johnson2020energy,johnson2021role} 
and thus be able to identify the small-scale mechanisms associated to local backscatter/inverse cascade events. And, further theoretically obtained exact relations between structure function and filtering approaches may yet be found.

\section*{Acknowledgements}
We thank Prof. Gregory Eyink for fruitful discussions and comments on this manuscript, Dr. Miguel Encinar for valuable suggestions, and the JHTDB/IDIES staff for their assistance with the database and its maintenance. This work is supported by the National Science Foundation (Grant \# CSSI-2103874). 

\section*{Declaration of interests}
The authors report no conflict of interest.

\appendix

\section{Turbulence database access tools}\label{appA}
The high-resolution isotropic DNS data are accessible via new python-based tools built upon the data housed in JHTDB (the Johns Hopkins Turbulence Database, \citep{li2008public}).  JHTDB has been operating for over a decade and has led to hundreds of peer-reviewed articles on turbulence. A new set of data access tools based on Jupyter Notebooks has been developed that enables direct access to subsets of the data continuing the “virtual sensors” concept \cite{li2008public}.
The new notebooks provide fast and stable operation on the existing turbulence data sets while enabling
user-programmable, server-side computations. To date, the new data access tools have been implemented and tested on the
high Reynolds number, forced isotropic turbulence data set on $8{,}192^3$ grid points (the {\it isotropic8192} datasets) of which 5 snapshots are at a Taylor microscope Reynolds number of $Re_{\lambda} = 1250$ \citep{yeung2012dissipation} and one with very high spatial resolution at $Re_{\lambda} = 610$.

The isotropic8192 data set has been partitioned into 4096 Zarr database files, each of which is a $512^3$ volume cubelet of the $8192^3$ data. Each Zarr file stores the velocity and pressure variables in distinct Zarr groups, and the data in each group is further broken down into chunks. The chunks are ijk-ordered such that cutouts and interpolation buckets, the size of which are dependent on the interpolation or differentiation method selected by the user, can be cutout directly from the intersecting chunk(s). 

The new Python-based data access tools, pyturb, are accessed via the SciServer (sciserver.org) platform. Pyturb interfaces directly with the data files in Zarr format, stored on volumes mounted to each user's SciServer container.  The entirety of the isotropic8192 data set ($8192^3$ volume, 6 snapshots) in Zarr format is publicly available through Python Notebook on SciServer. Users can apply for a SciServer account freely and download the demo Notebook.  In the Notebook, users can get access to pre-coded ``Get''  functions for arbitrary sets of points: {\it GetPressure} to retrieve and interpolate pressures, {\it GetPressureGradient} to retrieve and interpolate pressure gradient, and similarly {\it GetPressureHessian}, {\it GetVelocity}, {\it  GetVelocityGradient}, {\it GetVelocityHessian}, {\it GetVelocityLaplacian}, and {\it GetCutout} to read raw data for a user-specified box.

Demo codes for accessing data at user-specified arrays of points (in various sample geometrical configurations) are listed in the Notebook. The isotropic8192 datasets can be also accessed via the web-portal cutout service where the pyturb GetCutout function has replaced the legacy function for user queries (see \url{https://turbulence.pha.jhu.edu/newcutout.aspx}). JHTDB still provides and maintains other datasets (\url{https://turbulence.pha.jhu.edu/datasets.aspx}) through legacy SQL systems with C, Fortran, Matlab, Python, and .Net interface. However, the  aim is to transfer the existing datasets and any new coming datasets to the pyturb system in the future for faster and more stable services.

\section{Evaluating and comparing $ \langle k_{{\rm sf},\ell} \rangle$ and $ \langle k_{{\rm sgs},\ell} \rangle$}\label{appB}

The average values $\langle k_{{\rm sf},\ell} \rangle$ and $\langle k_{{\rm sgs},\ell} \rangle$ can be obtained from classical turbulence theory and the Kolmogorov spectrum. To evaluate $\langle k_{{\rm sf},\ell} \rangle$ we use the general expression for the structure function tensor in isotropic turbulence in the inertial range \cite{}
\begin{equation}
\langle \delta u _i({\bf r})\delta u _j({\bf r}) \rangle = 
C_2 \, \langle \epsilon \rangle^{2/3} \, r^{2/3}\left(
\frac{4}{3} \delta_{ij} - \frac{1}{3}\frac{r_i r_j}{r^2}.
\right),
\end{equation}
and write
\begin{equation}
\langle k_{{\rm sf},\ell} \rangle
=
\frac{1}{2 \,V_\ell}\iiint\limits_{V_{\ell}}  \frac{1}{2} \langle \delta u _i^2({\bf r}) \rangle d^3{\bf r}_s =
\frac{1}{4}\frac{1}{V_\ell}\iiint\limits_{V_{\ell}} 
 C_2 \, \langle \epsilon \, \rangle^{2/3} \, r^{2/3} \left(
\frac{4}{3} \delta_{ii} - \frac{1}{3}\frac{r_ir_i}{r^2}
\right) d^3{\bm r}_s
\end{equation}
where ${\bf r}_s = {\bf r}/2$. The integration yields
\begin{equation} 
\langle k_{{\rm sf},\ell} \rangle
=\frac{C_2 \, \langle \epsilon \rangle^{2/3} 
\ell^{2/3}}{4\times 3\pi(\ell/2)^3/4} 
\int\limits_{0}^{\ell/2}
 4 \pi r_s^2 \, (2r_s)^{2/3} \, dr_s = \frac{3}{4} \, C_2\, \langle \epsilon \rangle^{2/3} 
\ell^{2/3} \approx 1.575 \, \langle \epsilon \rangle^{2/3}
\ell^{2/3} 
\end{equation} 
when using the usual empirical Kolmogorov structure function constant $C_2 \approx 2.1$.

In order to evaluate $ \langle k_{{\rm sgs},\ell} \rangle$ we use   
\citep{pope_2000,li2004analysis}
\begin{equation}
\langle \tau_{ij} \rangle = \iiint \left( 1- \hat{G}^2_\ell({\bf k})\right) \, \Phi_{ij}({\bf k})\, d^3{\bf k} \,\, \,\,
\stackrel{\mathclap{\tiny (i=j)}}{=} 
\,\, \,\,
C_K \langle \epsilon \rangle^{2/3} \int_0^\infty \left( 1- \hat{G}^2_\ell({ k})\right) \, 2 \, k^{-5/3} d k
,
\label{eq:spectraltauii}
\end{equation}
where $\hat{G}_\ell({\bf k})=\hat{G}_\ell({k})$ is the Fourier transform of the filter function at scale $\ell$ and 
$\Phi_{ij}({\bf k}) = E(k)/(4\pi k^2)(\delta_{ij} - k_ik_j/k^2)$ is the spectral tensor for isotropic turbulence, while $E(k) = C_K \langle \epsilon \rangle^{2/3} 
k^{-5/3}$ is the radial 3D energy spectrum of turbulence. 

For the spherical top-hat filter, its Fourier transform can be shown to be
\begin{equation}
\hat{G}_\ell({k}) = \frac{3}{(k \ell/2)^3} \, \left(\sin \frac{k\ell}{2}  - \frac{k\ell}{2}  \cos \frac{k\ell}{2} \right)
\end{equation}
where $k=|{\bf k}|$.  The definite integral needed to evaluate the RHS of Eq. \ref{eq:spectraltauii} exists (using WolframAlpha online) and is given by 
\begin{equation}
\int \limits_0^\infty \left( 1-\left[ 
\frac{3}{(\kappa /2)^3} \, \left(\sin \frac{\kappa}{2}  - \frac{\kappa}{2}  \cos \frac{\kappa}{2} \right)  \right]^2 \right) \, \kappa^{-5/3} \, d\kappa = - 544 \,\, \Gamma(-20/3),
\end{equation}
with $\kappa = k \ell$ and where $\Gamma(..)$ is the Gamma function. Evaluating and using  $C_K \approx 1.6$, the result is
\begin{equation}
    \frac{1}{2}\tau_{ii} = 0.76 \, C_K
    \langle \epsilon \rangle^{2/3} \,\ell^{2/3} \approx 1.217 \, \langle \epsilon \rangle^{2/3} \,\ell^{2/3}.
\end{equation}

\bibliographystyle{jfm}
\bibliography{cascade}

\end{document}